\documentclass{aa}  

\usepackage{graphicx}
\usepackage{txfonts}
\usepackage{dsfont}
\usepackage{enumitem}
\usepackage[colorlinks=true,linkcolor=blue,filecolor=blue,citecolor=blue,urlcolor=blue]{hyperref}
\usepackage[normalem]{ulem}
\usepackage{caption}
\newcommand{\MAXI}{{MAXI~J1820+070}\xspace}
\newcommand{\NICER}{{\em NICER}\xspace}

\begin{document}

   \title{The nature of an imaginary quasi-periodic oscillation in the soft-to-hard transition of MAXI J1820+070}


   \author{Candela Bellavita
          \inst{1,2,3} 
          \and 
          Mariano M\'endez \inst{3}
          \and
          Federico Garc\'ia\inst{1,2}
          \and
          Ruican Ma\inst{4,5}
          \and
          Ole K\"onig\inst{6,7}
          }

   \institute{Instituto Argentino de Radioastronom\'ia (CCT La Plata, CONICET; CICPBA; UNLP), C.C.5, 1894 Villa Elisa, Argentina
   \and Facultad de Ciencias Astron\'omicas y Geof\'isicas, Universidad Nacional de La Plata, 1900 La Plata, Argentina
         \and
              Kapteyn Astronomical Institute, University of Groningen, PO BOX 800, NL-9700 AV Groningen, the Netherlands
          \and
              Key Laboratory of Particle Astrophysics, Institute of High Energy Physics, Chinese Academy of Sciences, Beĳing 100049, China
          \and
              Dongguan Neutron Science Center, 1 Zhongziyuan Road, Dongguan 523808, China
          \and
          Center for Astrophysics \textbar \ Harvard \& Smithsonian, 60 Garden Street, Cambridge, MA 02138, USA
          \and
              Dr. Karl Remeis-Observatory and Erlangen Centre for Astroparticle Physics, Friedrich-Alexander-Universität Erlangen-Nürnberg, Sternwartstr. 7, 96049 Bamberg, Germany
             }

   \date{Received xxx; accepted xxx}

 
  \abstract
  {A recent study shows that if the power spectra (PS) of accreting compact objects consist of a combination of Lorentzian functions that are coherent in different energy bands but incoherent with each other, the same is true for the Real and Imaginary parts of the cross spectrum (CS). Using this idea, we discovered imaginary quasi-periodic oscillations (QPOs) in NICER observations of the black hole candidate MAXI J1820+070. The imaginary QPOs appear as narrow features with a small Real and large Imaginary part in the CS but are not significantly detected in the PS when they overlap in frequency with other variability components. The coherence function drops and the phase lags increase abruptly at the frequency of the imaginary QPO. We show that the multi-Lorentzian model that fits the PS and CS of the source in two energy bands correctly reproduces the lags and the coherence, and that the narrow drop of the coherence is caused by the interaction of the imaginary QPO with other variability components. The imaginary QPO appears only in the decay of the outburst, during the transition from the high-soft to the low-hard state of MAXI J1820+070, and its frequency decreases from $\sim$5~Hz to $\sim1$~Hz as the source spectrum hardens. We also analysed the earlier observations of the transition, where no narrow features were seen, and we identified a QPO in the PS that appears to evolve into the imaginary QPO as the source hardens. As for the type-B and C QPOs in this source, the rms spectrum of the imaginary QPO increases with energy. The lags of the imaginary QPO are similar to those of the type-B and C QPOs above 2~keV but differ from the lags of those other QPOs below that energy. While the properties of this imaginary QPO resemble those of type-C QPOs, we cannot rule out that it is a new type of QPO.
}

   \maketitle
%

\section{Introduction} \label{sec:intro}

Black hole X-ray binaries (BHXBs), consisting of a black hole and a secondary star, exhibit X-ray outbursts driven by mass accretion onto the black hole \citep{Bahramian2023}. As they transition through different X-ray spectral states \citep{Mendez1997,Belloni2005}, BHXBs describe a “\textit{q}” shape in the hardness intensity diagram \citep[HID;][]{Homan2001}. In a typical outburst, a transient BHXB starts in the low hard state (LHS), with a relatively low X-ray intensity and a hard energy spectrum \citep{Mendez1997,Belloni2005}. During this state, the source follows an almost vertical path in the diagram as the intensity increases and the hardness ratio decreases slightly \citep{Belloni2011}. Eventually, the source spectrum softens at an essentially constant intensity, and the BHXB transitions to the hard and soft intermediate states \citep[HIMS and SIMS, respectively;][]{Homan2005}. As the outburst continues, the source then reaches the high soft state (HSS) where the energy spectrum is soft and the intensity decreases while its hardness remains more or less constant \citep{Mendez1997,Belloni2005,Belloni2011}. Finally, at the decay of the outburst, the BHXB returns to the LHS, completing the \textit{q} shape in the diagram \citep{Belloni2010}.

BHXBs exhibit fast X-ray variability throughout outbursts, offering crucial insights into the dynamics of the accretion processes around the compact object \citep{vanderklis1994,vanderKlis2006, Belloni2011}. The analysis of this variability is key to understanding the mechanisms of accretion and ejection involved in these systems \citep[][see \citealt{Motta2016} for a review]{vanderklis1994,Fender2004, Ingram2009,Kara2019,Mastroserio2019}. The power spectrum (PS) of these sources presents relatively narrow peaks called quasi-periodic oscillations (QPOs), which are among the most notable features of the variability \citep[][see \citealt{IngramMotta2019} for a review]{vanderklis1989,Nowak2000,Belloni2002}. Low-frequency QPOs (LFQPOs) in BHXBs have frequencies ranging from a few mHz to $\sim 30$~Hz \citep{Belloni2002,Casella2004,Remillard2006,Motta2012}. LFQPOs are classified into three types \citep[][for reviews, see \citealt{Motta2016,Belloni2016}]{Wijnands1999, Remillard2002, Casella2004,Casella2005}, A, B and C, based on their centroid frequency, $\nu_0$, quality factor $Q=(\nu_0/\rm FWHM)$, where FWHM is the full width half maximum of the QPO, fractional root mean square (rms) amplitude, phase lags, and the strength and shape of the broadband noise in the PS \citep{Casella2004}. Among these, type-C QPOs are the most common in BHXBs \citep{Motta2012}. These QPOs are strong, with rms amplitudes that can reach up to 20\%, and are generally observed in the LHS and the HIMS \citep{Casella2004,Motta2011}. Type-C QPOs have centroid frequencies in the range of $0.1-15$~Hz and high quality factors, normally above 10 \citep{Casella2004,Belloni2005,Motta2011}. On the other hand, type-B QPOs are weaker, with rms amplitudes lower than 5\%, and centroid frequencies around 6~Hz \citep{Casella2004,Casella2005}. These QPOs appear only in the SIMS, distinguishing this state from the HIMS \citep{Motta2011}. Type-B QPOs usually have quality factors above 6 and are accompanied by weak red noise \citep{Casella2004,Casella2005}. Type-A QPOs are the weakest LFQPOs and they usually appear just after the source transitions into the HSS with frequencies around $6-8$~Hz \citep{Belloni2014}.

The Fourier cross-spectrum (CS) between simultaneous light curves in two energy bands is employed to derive the frequency-dependent phase lags between those two energy bands \citep{vanderklis1987,Vaughan1998,Nowak1999,Uttley2014}, which give the phase angle of the cross vector in the complex Fourier plane \citep{Vaughan1997}. Furthermore, the CS and the PS of the two time series can be used to obtain the coherence function. The coherence function is a measure of the degree of linear correlation between the two simultaneous time series as a function of the Fourier frequency \citep{bendat2011random,Vaughan1997}, and allows us to identify a particular frequency or range of frequencies that are strongly correlated across energy bands \citep{Nowak1999}. The coherence function is a valuable tool for studying processes characterised by features such as QPOs or abrupt changes in slope in the PS \citep{Nowak1999}. \cite{Vaughan1997} showed that if there are multiple signal components contributing to the data in two energy bands, the coherence function may fall below unity, even if each individual component generates perfectly coherent variability.

\cite{Mendez2024} presented a novel method to measure the lags of variability components in low-mass X-ray Binaries (LMXBs). Employing a multi-Lorentzian model --a linear combination of Lorentzian functions-- the authors fitted simultaneously the PS and the Real and Imaginary parts of the CS. \cite{Mendez2024} assumed that both the PS and the CS comprise several components that are coherent in different energy bands but incoherent with each other. In that case, the Real and Imaginary parts of the CS are linear combinations of the same Lorentzians but each of them is multiplied by, respectively, the cosine or the sine of a, in principle frequency-dependent, function that represents the phase lag between the corresponding Lorentzian in each signal. Once they find the best fit for the PS and the CS, they can derive the model that predicts the lags and coherence function. 

The technique introduced by \citet{Mendez2024} allowed them to unveil variability components that were not significantly detected in the PS but were significant in the CS. Using this method, the authors found a narrow QPO in an observation of the BHXB \MAXI that was very significant in the Imaginary part of the CS but was not detected in the PS. They referred to this QPO as “imaginary QPO” since it has a large Imaginary but a small Real part in the CS. This signal overlaps in frequency with other signals with large Real parts, which leads to a narrow drop in the coherence function at the imaginary QPO frequency \citep[first observed by][see also \citealt{Ji2003}]{Konig2024}. 

\MAXI is an X-ray transient discovered with the Monitor of All-Sky X-ray Image \citep[MAXI;][]{Matsuoka2009} in 2018 \citep{Kawamuro2018}, when the source underwent an outburst. \MAXI was closely monitored by the \textit{Neutron Star Interior Composition Explorer} \citep[\NICER;][]{Gendreau2016}, almost daily from March 6 to November 21; within these dates, the source exhibited significant spectral changes, describing an overall \textit{q} shape in the HID. During the transition from the LHS to the HSS, a type-C QPO present in the X-ray power of \MAXI switched to a type-B QPO \citep{Homan2020}. \cite{Ma2023} studied the observation that covered the transition of the QPO types (obsID~1200120297) and obtained the rms and phase lag spectra of both types of QPOs.  \cite{Ma2023} found that above $1.5-2.0$~keV, the rms and lag spectra of the two types of QPOs are similar, but they are significantly different at lower energies: below $1.5-2.0$~keV, the type-B QPOs show a smaller rms amplitude and softer phase lags than the type-C QPOs. 

The purpose of this work is to expand the analysis carried out by \cite{Mendez2024}, searching for other observations of \MAXI that exhibit the drop in the coherence to study the properties of the imaginary QPO. In Sec.~\ref{sec:math} we show why a variability component with a large Imaginary part in the CS, when overlapping in frequency with a component with a large Real part, can be undetectable in the PS  but significantly detected in the CS. In Sec.~\ref{sec:obs} we describe the observations and data reduction of \MAXI. In Sec.~\ref{sec:results} we present the results obtained from the analysis of the \MAXI data that exhibit a drop in the coherence. We then fit those observations and we study the properties of the imaginary QPO. Finally, in Sec.~\ref{sec:discussion} we discuss our findings. 


%

\section{Hidden variability} \label{sec:math}
The novel method presented by \citet{Mendez2024} allows us to detect variability components that are not detected in the PS but are significant in the CS. As demonstrated in the following paragraphs, this happens when a signal that has a large Imaginary part but a small Real part in the CS overlaps in frequency with other signals that have a large Real part. 

Consider two correlated light curves of an LMXB measured in two energy bands, denoted as $x(t)$ and $y(t)$, with corresponding complex Fourier transforms $X(\nu)$ and $Y(\nu)$. Following the notation in \citet{Mendez2024}, we define the PS of both series as $G_{xx}(\nu)$ and $G_{yy}(\nu)$, the CS as $G_{xy}(\nu)=\langle |X(\nu)||Y(\nu)| e^{i\Delta\phi_{xy}(\nu)}\rangle$, where $\Delta\phi_{xy}(\nu)$ is the phase lag between the two series at frequency $\nu$, and the angle brackets indicate averaging over an ensemble of measurements of $X(\nu)$ and $Y(\nu)$ \citep[for details, see][]{bendat2011random}. We define the intrinsic coherence function as $\gamma^2_{xy}(\nu)=|G_{xy;i}(\nu)|^2/{G_{xx}(\nu) G_{yy}(\nu)}$. We will consider two variability components overlapping in frequency, one with a phase lag of $\pi/2$, such that its Real part in the CS, $\rm Re_1(\nu)$, is zero, and the other one with a phase lag of $0$, such that its Imaginary part, $\rm Im_2(\nu)$, is zero. We describe each variability component with a Lorentzian function, coherent in both energy bands but incoherent with each other. We assume that the variability components are additive \citep{Nowak2000, Belloni2002, Mendez2024}, but there are other possibilities \citep[see for instance][]{Ji2003,Uttley2005, Ingram2013, Zhou2022}. We can write: 
\begin{equation}
\begin{aligned}
    G_{xx;i}~(\nu) = A_i L(\nu;\nu_{0;i},\Delta_i) \\
    G_{yy;i}~(\nu) = B_i L(\nu;\nu_{0;i},\Delta_i), 
\end{aligned}
\label{eq:PSs}
\end{equation}
where $i=1,2$ refers to the $i$th variability component, $A_i, B_i \in \mathds{R}$ are the integrated power, from zero to infinity, of the Lorentzian components in each of the two energy bands and $\nu_{0;i}$ and $\Delta_i$ are, respectively, the centroid frequency and the FWHM of the Lorentzians. These two Lorentzian parameters, $\nu_{0;i}$ and $\Delta_i$, are the same in both energy bands, assuming that each input process is perfectly coherent with the corresponding output process. 

Under the assumption that for each Lorentzian function $\gamma^2_{xy;i}(\nu)=1$, $|G_{xy;i}(\nu)|^2= A_i B_i L^2(\nu;\nu_{0;i},\Delta_i)$, and from the choice that the first component has a zero Real ($\Delta\phi_{xy;1}(\nu)=\pi/2$) part and the second component a zero Imaginary part ($\Delta\phi_{xy;2}(\nu)=0$):
\begin{equation}
    \begin{aligned}
        {\rm Im}[G_{xy;1}]= \sqrt{A_1 B_1} \\
        {\rm Re}[G_{xy;2}]= \sqrt{A_2 B_2},
    \end{aligned}
    \label{eq:im&re}
\end{equation}
where $G_{xy;i} = \int_0^{\infty} G_{xy;i}(\nu)d\nu$.

Let us posit two additional assumptions regarding the considered variability components: 
\begin{enumerate}[label=(\roman*)]
\item ${\rm Component}~2$ is stronger than ${\rm Component}~1$: ${\rm Re}[G_{xy;2}] \gg {\rm Im}[G_{xy;1}]$ and therefore, $A_2B_2 \gg A_1B_1$
\item The rms spectrum of both components is similar, such that $B_2/A_2=B_1/A_1=:k$. 
\end{enumerate}
Combining these two hypotheses, we have that $A_2 \gg A_1$ and $B_2 \gg B_1$, and we can approximate the total PS as:
\begin{equation}
    \begin{aligned}
        G_{xx}=G_{xx;1}+G_{xx;2}= A_1+A_2 \approx A_2 \\
        G_{yy}=G_{yy;1}+G_{yy;2}= B_1+B_2 \approx B_2.
    \end{aligned}
\end{equation}
The error for the Imaginary part of the CS \citep{bendat2011random,Ingram2019}, is:
\begin{equation}
        {\rm dIm} [G_{xy}] = \sqrt{\frac{G_{xx} G_{yy}-{\rm Re}^2[G_{xy;2}] + {\rm Im}^2[G_{xy;1}]}{2N}} \approx \sqrt{\frac{A_1B_1}{2N}},
\label{eq:bendat}
\end{equation}
where $N$ is the product of the number of segments used to compute the CS 
and the number of frequency bins in the range in which we measure the CS. Hence, the signal-to-noise ratio (SNR) of the ${\rm Component}~1$ in the CS is:
\begin{equation}
    {\rm SNR}_{xy;1} \approx \frac{\sqrt{A_1B_1}}{\sqrt{A_1B_1/(2N)}}=\sqrt{2N}.
\label{eq:SNRcs}
\end{equation}
As a consequence, given a sufficient number of segments, ${\rm Component}~1$ will attain significance in the imaginary part of the CS. 

For the PS in each energy band, using (ii), we can write Eq. \ref{eq:im&re} as ${\rm Im} [G_{xy;1}]= \sqrt{k}A_1$ and ${\rm Re} [G_{xy;2}]= \sqrt{k}A_2$, and it is easy to show that:
\begin{equation}
    {\rm d}G_{xx} \approx \frac{{\rm Re} [G_{xy;2}]}{\sqrt{kN}}= \frac{A_2}{\sqrt{N}} ~~~~~~~~;~~~~~~~~ {\rm d}G_{yy} \approx \frac{B_2}{\sqrt{N}}.
\end{equation}

Hence, the SNR of the ${\rm Component}~1$ in each PS is:
\begin{equation}
    {\rm SNR}_{xx;1} \approx A_1 \frac{\sqrt{N}}{A_2} ~~~~~~~~;~~~~~~~~ {\rm SNR}_{yy;1}=B_1 \frac{\sqrt{N}}{B_2}.
\label{eq:SNRps}
\end{equation}
Subsequently, since $A_2 \gg A_1$ and $B_2 \gg B_1$, the SNR of the ${\rm Component}~1$ increases with $N$ much more slowly in the PS than in the CS.  

We have demonstrated that a QPO characterised by a large Imaginary part and a small Real part in the CS can be hidden in the PS, while being significant in the CS. We emphasise \citet{Mendez2024} conclusion that searching for QPOs exclusively in the PS may lead to overlooking significant variability components. In the next sections, we will use these considerations to study the variability of \MAXI. 

As shown in \citet{bendat2011random}, the error of the coherence function is ${\rm d}\gamma^2_{xy} = {\sqrt{2}(1-\gamma^2_{xy})|\gamma_{xy}|}/{\sqrt{N}}$. Therefore, when the coherence function is close to unity, its estimates can be more accurate than those of the PS or CS. Consequently, variability components hidden in both PS and CS could have a large SNR in the coherence function. Given its greater sensitivity, the coherence function may allow us to uncover new signals that could be missed by looking only at the PS and CS.

\section{Observations and data analysis} \label{sec:obs}

We use the tool \textit{nicerl2} to process the \NICER data of \MAXI and produce the clean event files. We compute the PS in two energy bands, $0.3-2.0$~keV and $2.0-12.0$~keV, and the CS between the same bands using GHATS\footnote{\url{http://astrosat-ssc.iucaa.in/uploads/ghats\_home.html}}. We calculate the Fast Fourier transform (FFT) in each band, setting the length of the segment to 65.536~s and the time resolution to 0.4~ms, which results in a lowest frequency of 0.015~Hz and a Nyquist frequency of 1000~Hz. GHATS generates the Leahy normalised PS and CS for each segment, which are then averaged to produce the PS and CS of the observation. To correct for the Poisson level in the PS, we subtract the average power in the frequency range $400-800$~Hz. To increase the SNR at intermediate and high frequencies, we perform a logarithmic rebinning, increasing the size of each bin by a factor $10^{1/100}$ compared to the previous bin. We finally normalise the PS and CS to fractional rms units, ignoring the background since its contribution to the total count rate is negligible. GHATS also computes the phase-lag and the coherence-function frequency spectra. We take the energy band $0.3-2.0$~keV as the reference band when computing the CS and phase-lag frequency spectrum. 

We note that the phase lags of \MAXI during these observations are close to zero over a broad range of frequencies, meaning that the Imaginary part of the CS is much smaller than the Real part at all Fourier frequencies. We therefore rotate all cross vectors by 45 degrees to have approximately equal Real and Imaginary parts. As explained in \cite{Mendez2024}, this rotation would make an eventual fit of the CS more stable without having any impact on the best-fitting parameters. 

We process the data of 131 \NICER observations, from obsID 1200120101 to 1200120293, that correspond to MJD $58190-58443$. We compute the light curve of \MAXI for each observation in the $0.5-12.0$~keV band. We use \textit{nicer3-lc} to normalise the light curves by the number of active focal plane modules (FPMs) that, in some cases, had to be lowered to avoid data dropouts caused by saturation when the source was too bright. Defining the hardness ratio as the ratio of the count rates in the $2.0-12.0$~keV to that in the $0.5-2.0$~keV, we produce the hardness intensity diagram (HID). 

We explore the phase lags and coherence function for all the observations, aiming to identify sharp features that could result from the overlapping in frequency of two or more variability components. For the observations in which this is the case, we repeat the procedure to generate the PS and CS but using two hard energy bands, $2.0-5.0$~keV and $5.0-12.0$~keV, and we examine if those features remain visible. 

We use \texttt{XSPEC v.12.13.0c} to fit simultaneously the PS in the $0.3-2.0$~keV and $2.0-12.0$~keV bands and the Real and Imaginary parts of the CS in the same energy bands, considering the $0.01-50$~Hz frequency range. We begin by fitting simultaneously the PS in both bands with a single Lorentzian function and we add Lorentzians until the reduced $\chi^2$ is about unity and there are no systematic trends in the residuals. The multi-Lorentzian model obtained from this phenomenological approach serves as a starting point to fit simultaneously both PS and the Real and Imaginary parts of the CS assuming constant phase lags with Fourier frequency for each component \citep{Mendez2024}. We link the centroid frequencies and FWHM of each Lorentzian in the two PS and in the CS but we leave the normalisations of the PS free to vary independently. For each Lorentzian, its normalisation in the CS is tied to the square root of the product between the two PS normalisations. If necessary, we add more Lorentzians until we obtain a good fit, now considering also the residuals of the CS. 

From the best-fitting model, we derive the model for the phase lags and the coherence function. While we plot the phase-lag spectrum and the coherence function with the derived model, it is important to note that we do not fit those data; rather those models are a prediction of the multi-Lorentzian model used to fit PS and CS.

To explore the energy dependence of the rms amplitude and phase lags of a QPO, we extract the PS in six energy bands: $0.3-1.0$, $1.0-1.5$, $1.5-2.5$, $2.5-4.0$, $4.0-5.0$, $5.0-12.0$~keV. We also generate the CS of each band with respect to the total band $0.3-12$~keV, which is therefore the reference band for the lags. As usual, we refer to hard lags when the high-energy photons lag the low-energy ones, and to soft lags when the opposite occurs. To correct for the partial correlation of the photons that are simultaneously in the narrow and the total energy bands, we subtract the average of the Real part of the CS calculated in a frequency range where the source does not contribute \citep{Belloni2024,Mendez2024}. Using the best-fitting model for each observation with the centroid frequencies and FWHMs of every Lorentzian fixed, we fit simultaneously the PS in a small band of energy, the PS in the total band and the Real and Imaginary parts of the CS. We then construct the amplitude and phase-lag spectra using the parameter values of the Lorentzian corresponding to the QPO in each small energy band: the rms amplitude at that energy band is the square root of the Lorentzian normalisation, and the phase lag is the argument of the cosine and the sine functions that multiply the Lorentzian in, respectively, the Real and Imaginary parts of the CS.

\section{Results} \label{sec:results}

\begin{figure*}
    \centering
    \includegraphics[width=\textwidth]{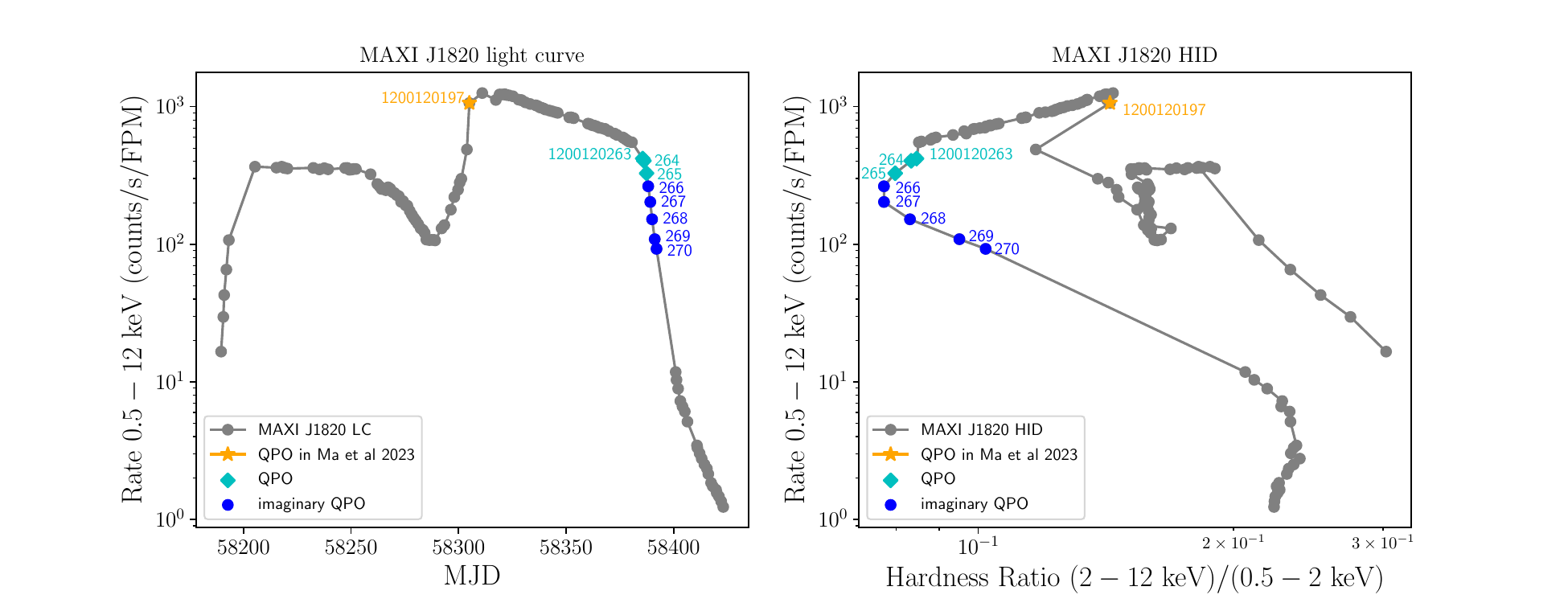}
    \caption{\NICER X-ray light curve (left panel) and hardness-intensity diagram (right panel) of \MAXI during its 2018 outburst. The intensity in both panels is the count rate per detector in the $0.5-12$~keV band, while the hardness ratio in the right panel is the ratio of the count rates in the $2-12$~keV to that in $0.5-12$~keV. Each point corresponds to one obsID within $1200120101-1200120293$. Light blue diamonds (obsIDs $1200120263-1200120265$) and blue dots (obsIDs $1200120266-1200120270$) depict the observations considered in this paper, in particular, the second ones correspond to those presenting narrow features in the phase lags and the coherence function. The orange star represents the observation 1200120197 studied by \cite{Ma2023}, which covered the transition of type-C to a type-B QPO.}
    \label{fig:HID}
\end{figure*}

The left panel of Fig.~\ref{fig:HID} shows the evolution of the \NICER count rate of the source in the $0.5-12$~keV band during the outburst. Each data point represents one obsID from 1200120101 to 1200120293, which corresponds to MJDs between 58189 and 58424. Throughout this period, the source traced an overall \textit{q} shape in the HID, as shown in the right panel of Fig.~\ref{fig:HID}. At the beginning of the outburst, \MAXI is in the LHS and its count rate increases rapidly from  $\sim$16~$\rm{~cts~s^{-1}~FPM^{-1}}$ to $\sim$360~$\rm{~cts~s^{-1}~FPM^{-1}}$ as the source transitions to the intermediate state. At that point, the count rate stays more or less constant, except for a small valley between MJD $58262-58305$ that reaches a minimum of $\sim$106~$\rm{~cts~s^{-1}~FPM^{-1}}$. The source then enters the HSS, where the count rate decreases slowly from $\sim$1200~$\rm{~cts~s^{-1}~FPM^{-1}}$ to $\sim$~550~$\rm{~cts~s^{-1}~FPM^{-1}}$. Around MJD~58380, \MAXI leaves the HSS and the count rate drops rapidly as the source returns to the LHS, reaching $\sim$1.2~$\rm{~cts~s^{-1}~FPM^{-1}}$ by the last observation considered. Days after approaching quiescence \citep{Russell2019}, \MAXI underwent three reflares \citep{Ulowetz2019,Hambsch2019,Adachi2020}, but we do not analyse them in this study.

\begin{figure*}
    \centering
    \includegraphics[width=0.9\textwidth]{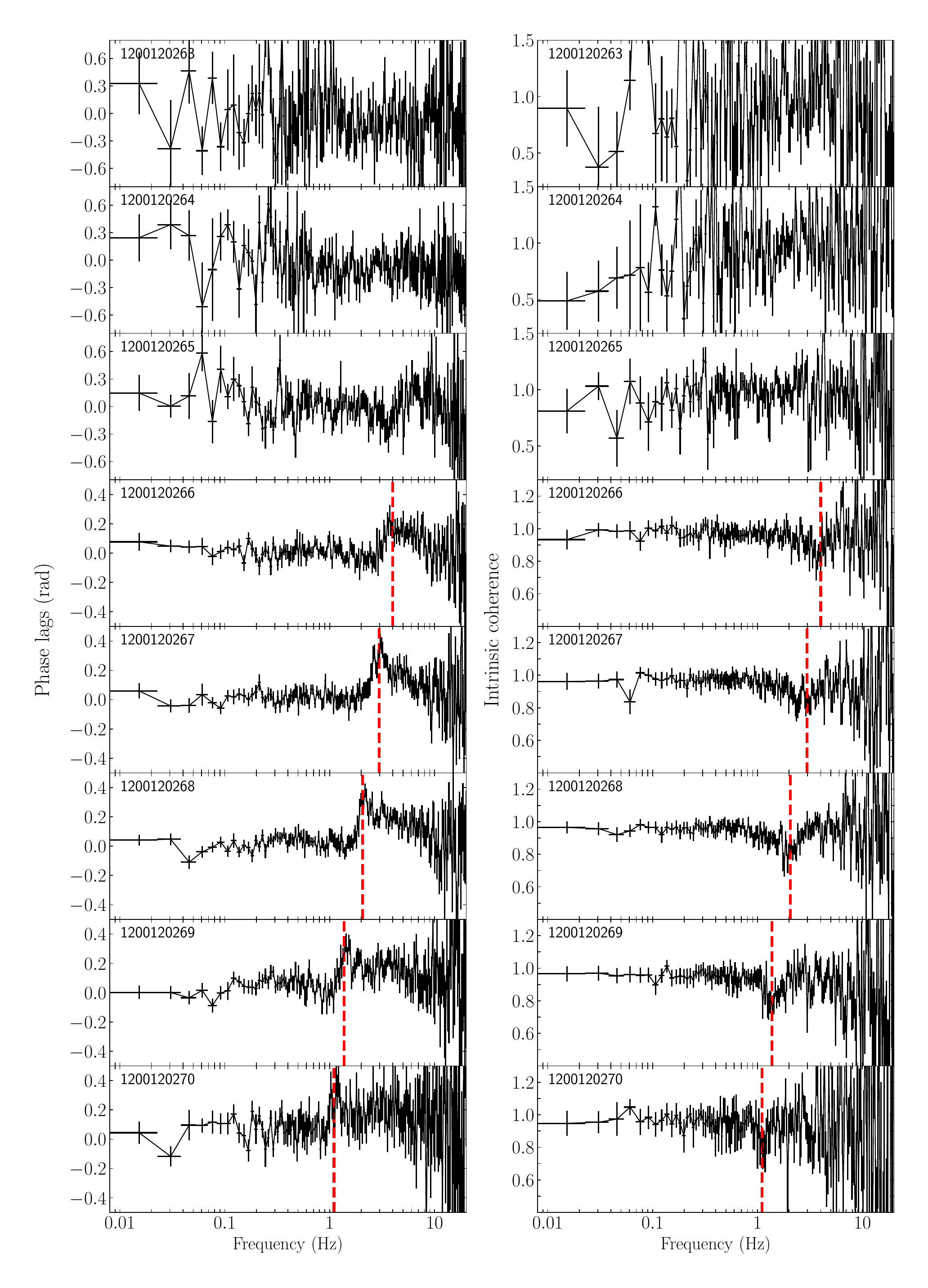}
    \caption{Phase lags (left panels) and intrinsic coherence function (right panels) for \NICER observations from 1200120263 to 120012070 of \MAXI for the 0.3$-$2~keV and 2$-$12~keV energy bands. With a red dashed line, we depict the frequency of the imaginary QPO in each observation derived from the multi-Lorentzian fitting approach (see Sec.~\ref{sec:imagQPO} for details)}
    \label{fig:lagycohe}
\end{figure*}

We examine the phase lags and coherence function of all 131 observations searching for significant features, such as sharp drops or abrupt changes, which may indicate overlapping variability components and potentially unveil a hidden component in the PS. We identify some observations in the HIMS at the upper part of the $q$ where a shallow drop in the coherence coincides with a QPO in the PS. Furthermore, we note that, during five observations in the lower branch of the $q$, where the source is transitioning back to the LHS, the coherence exhibits a more pronounced and significant drop at a frequency at which, in principle, no QPO is visible in the PS. At the same frequency of this coherence feature, we observe an abrupt jump in the phase lags that resembles a `cliff'. These observations provide an excellent opportunity to apply the derived model presented in \cite{Mendez2024} and assess its predictive capability. Therefore, aiming to identify a variability component hidden in the PS that can explain these significant features, we focus on these five observations, which are highlighted in blue in Fig.~\ref{fig:HID} and correspond to ObsIDs from 1200120266 to 1200120270.

To encompass the complete HSS-to-LHS transition, we extend the analysis to include the obsIDs 1200120263, 1200120264 and 1200120265, depicted with light blue diamonds in Fig.~\ref{fig:HID}. Before observation 1200120263, \MAXI was in the HSS, where the rms amplitude of the variability is very low.  After observation 1200120270, there was a nine-day gap in the data. In the following observation, the source was already in the LHS with a count rate almost an order of magnitude lower, and therefore much weaker.

In Fig.~\ref{fig:lagycohe} we show the phase lags (left panels) and the coherence function (right panels) for each observation of the HSS-to-LHS transition. In the first three observations, no significant features are visible. For the following observations, in the left panels, we see that the phase lags increase more or less steeply at the frequency marked by the red dashed line, and then decrease more or less smoothly as the frequency increases, resembling a cliff; the cliff moves to lower frequencies as the spectrum of \MAXI hardens. Concomitantly, in the right panels, we see a sharp drop in the coherence function that moves to lower frequencies as the source approaches the LHS. 

In particular, during observations 1200120264 and 1200120266, the source exhibited a significant change in the count rate. We examine the dynamical power spectra and observe notable differences in the power behaviour between the two periods. Therefore, for the subsequent analysis, we divide those observations into two segments, named p1 and p2. For observation 1200120264, the segments correspond to orbits 1 and 2-4, and for observation 1200120266, they correspond to orbits 1 and 2-5. For convenience, we will hereafter refer to each of these segments as an `observation'.

\subsection{Imaginary QPO}
\label{sec:imagQPO}
In this section, we present the results of fitting the \NICER observations during the HSS-to-LHS transition of \MAXI. As explained in Sec.~\ref{sec:obs}, for each observation we start by simultaneously fitting a multi-Lorentzian model to the PS in two energy bands, $0.3-2.0$~keV and $2.0-12.0$~keV, considering almost 4 frequency decades, from $0.01$~Hz to $50$~Hz. We then use this model as a starting point to fit both PS and the Real and Imaginary parts of the CS following the method introduced by \citet{Mendez2024} assuming constant phase lags. In every case, we add the number of Lorentzian functions needed to obtain a reduced $\chi^2$ near 1 and residuals without trends. In this phenomenological approach, all of the Lorentzian components have a significance of at least $3-\sigma$ in either the PS, the CS or both, which we measure from the error (the negative one if the errors are asymmetric) of the normalisation of the Lorentzian.  In the first observation, we fit a model with four Lorentzians; for the following two observations, a model with three; for the next six observations, a model with five; and for the last one, a model of six Lorentzians provides a good fit. In Appendix~\ref{sec:appendix}, we present the best-fitting models for each observation (see Fig.~\ref{fig:allfits}), as well as the best-fitting parameters with their $1-\sigma$ uncertainties (see Table~\ref{tab:longtable}). We use the derived model explained in \cite{Mendez2024} to predict the behaviour of the phase lags and the coherence function. 

\begin{figure*}
    \centering
    \includegraphics[width=\textwidth]{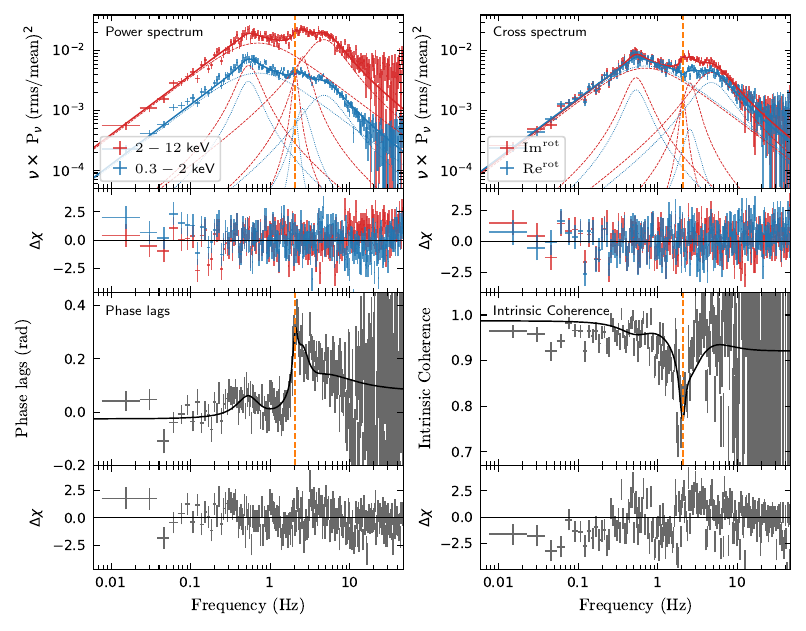}
    \caption{Top left panel: PS of \NICER observation 1200120268 of \MAXI in two energy bands. The 0.3$-$2~keV data is shown in blue while the 2$-$12~keV one is in red, both with the best-fitting model (solid line) consisting of 5 Lorentzian functions (dotted lines). Top right panel: Real and Imaginary parts of the CS between the same two energy bands rotated by $45^{\circ}$. We plot $\rm{Re}~cos(\pi/4) -\rm{Im}~sin(\pi/4)$ in blue and $\rm{Re}~sin(\pi/4) +\rm{Im}~cos(\pi/4)$ in red, with the best-fitting model assuming constant phase lags. Bottom panels: Phase-lag spectrum (left) and intrinsic coherence function (right) with the derived model (solid line) obtained from the fit to the PS and CS. In all four panels, residuals with respect to the model, defined as $\Delta \chi = {\rm (data-model)/error}$, are also plotted. The orange dashed vertical line at $2.06$~Hz in every panel corresponds to the centroid frequency of the imaginary QPO in the model.}
    \label{fig:fit268}
\end{figure*}

We take observation 1200120268 as an example to show our findings. The best fit of these data with a five-Lorentzian model gives a $\chi^2 \approx 992$ for 823 d.o.f. In the top panels of Fig.~\ref{fig:fit268} we present the best fit to the two PS (left) and the Real and Imaginary parts of the CS (right), with their corresponding residuals. In the bottom panels of Fig.~\ref{fig:fit268} we show the prediction of the derived model for the phase lags (left) and the coherence function (right). We note that the behaviour of both of them is well-reproduced by the model. Around 2.06~Hz, the coherence drops significantly, while the phase lags become abruptly harder. We plot a dashed vertical line at $2.06$~Hz in every panel, corresponding to the centroid frequency of one of the QPOs in the model. We refer to this QPO as the imaginary QPO, since it has a large imaginary part and a small real part in the CS. While this imaginary QPO is not significant in either of the PS, it is needed in the CS and it is responsible for the narrow drop in the coherence function.

Previous studies have fitted the PS of \MAXI with two or three broad components \citep[e.g.,][]{Kawamura2023}. This may appear to contradict the need to use up to five Lorentzians to fit the PS and the CS simultaneously. For instance, \cite{Veledina2016} suggested that peaked noise observed in the PS of accreting BHXBs could be the result of the interference of two broad components, the disc Comptonization and the synchrotron Comptonization, both modulated by accretion rate fluctuations and separated by a time delay. To explore this idea, we attempt to fit the data with only two broad Lorentzians and assess whether such a model can explain the observed drop in the coherence. In doing this, we obtain structured residuals in the PS, CS, phase lags and coherence. In particular, the derived model for the coherence shows a broad and shallow drop, and at around 2~Hz the coherence shows narrow residuals at a $\sim$6$\sigma$ level. In summary, a model with only two broad Lorentzians cannot fit the data. Specifically, this model does not reproduce the drop in the coherence at $\sim$2.1~Hz. The only way to recover the narrow and deep drop observed in the coherence data is to include the imaginary QPO.  

Since \cite{Konig2024} remarked that the feature in the coherence function is only observed when an energy band below 2~keV is used, we also extract the PS in two hard energy bands, 2$-$5~keV and 5$-$12~keV, and the CS between these two energy bands for the observations considered in this work. We notice that the sharp features shown in Fig.~\ref{fig:lagycohe}, the narrow drop in the coherence and the cliff in the phase lags, are no longer detected in this case. Using the best-fitting model obtained for the energy bands 0.3$-$2~keV and 2$-$12~keV, we fit the data in these hard bands, fixing the centroid frequencies and FWHM of every Lorentzian, and we find that the imaginary QPO is not significant in any of the PS neither in the CS. The $95\%$ confidence upper limit to the rms fractional amplitude of that Lorentzian in the PS corresponding to 5$-$12~keV is $\sim$4\% and in the 2$-$5~keV and in the CS is $\sim$3\%. 

\begin{figure*}
    \centering
    \includegraphics[width=\textwidth]{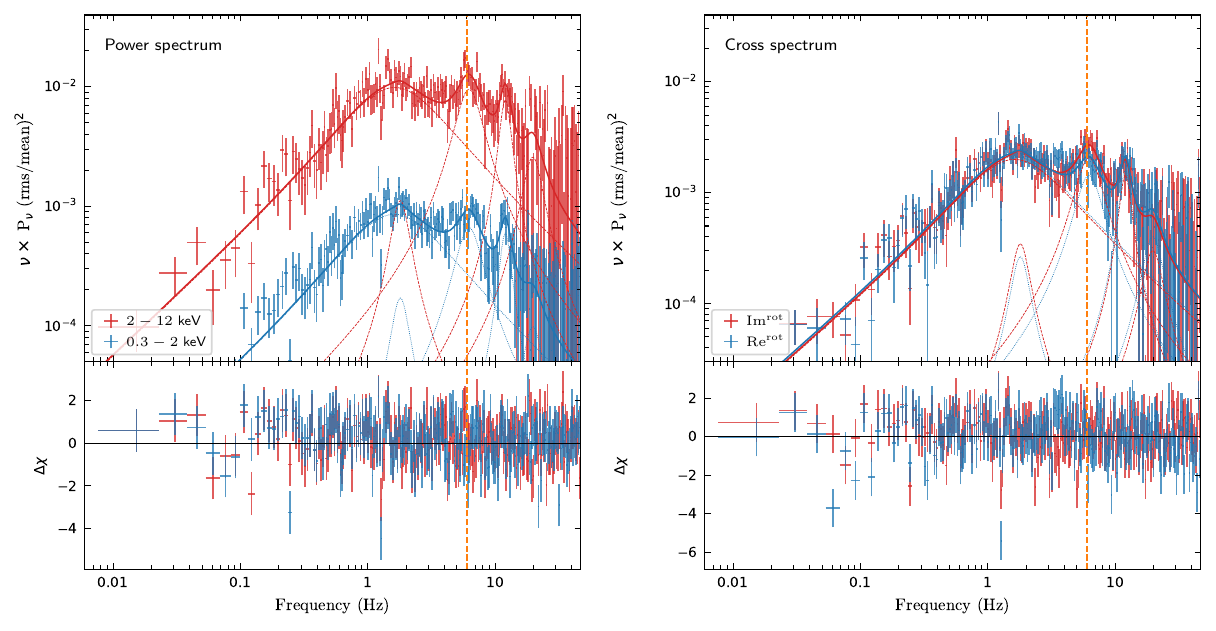}
    \caption{Left panel: PS of \NICER observation 1200120265 of \MAXI in two energy bands. The $0.3-2$~keV data is shown in blue while the $2-12$~keV one is in red, both with the best-fitting model (solid line) consisting of 5 Lorentzian functions (dotted lines). Right panel: Real and Imaginary parts of the CS between the same two energy bands rotated by $45^{\circ}$. We plot $\rm{Re}~cos(\pi/4) -\rm{Im}~sin(\pi/4)$ in blue and $\rm{Re}~sin(\pi/4) +\rm{Im}~cos(\pi/4)$ in red, with the best-fitting model assuming constant phase lags. In both panels, residuals with respect to the model, defined as $\Delta \chi = {\rm (data-model)/error}$, are also plotted. The orange dashed vertical line at $6.03$~Hz in each panel corresponds to the centroid frequency of the QPO in the model.}
    \label{fig:fit265}
\end{figure*}

   \begin{figure*}
   \centering
   \includegraphics[width=\textwidth]{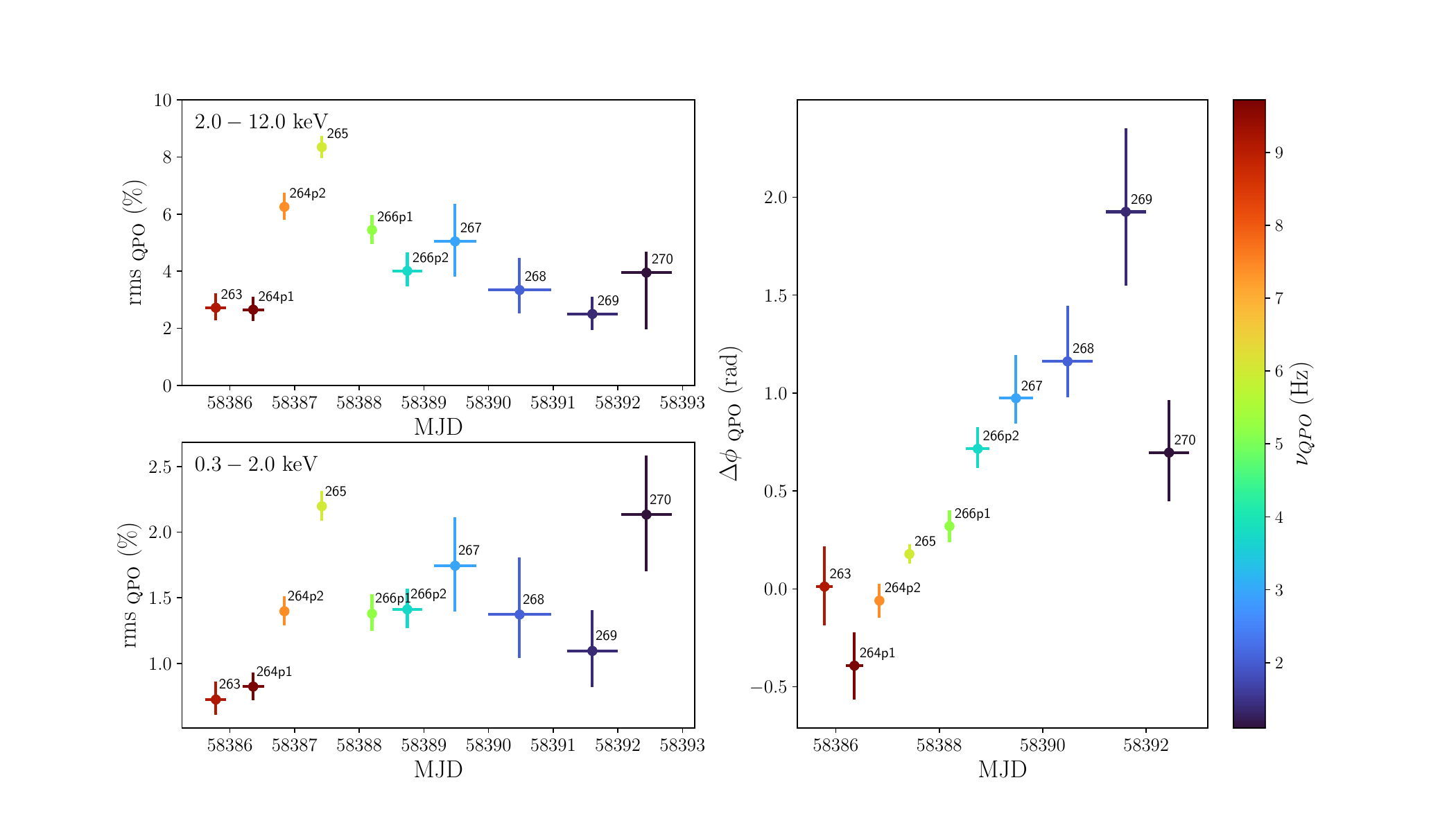}
   \caption{Evolution of the rms amplitude of the QPO in the $2.0-12.0$~keV band (top left) and in the $0.3-2.0$~keV band (bottom left) as \MAXI transitions out of the HSS. The right panel displays the evolution of the phase lags of the QPO for the $2.0-12.0$~keV band with respect to the $0.3-2.0$~keV band. The rms and phase lags are derived from the fitted Lorentzian in our model (see Sec.~\ref{sec:obs}). In the three panels, the horizontal bars correspond to the time duration of each observation, while the vertical bars are the actual $1-\sigma$ errors of each quantity. The colour gradient depicts the centroid frequency of the QPO.}
              \label{fig:rmsylagvst}%
    \end{figure*}

We use the same method described above to fit all the observations of the HSS-to-LHS transition of \MAXI. Moving backwards in time from observation 1200120268, we also identify the imaginary QPO that naturally explains the narrow features in the phase lags and coherence function. As in \cite{Mendez2024}, we find that the peak of the cliff in the phase lags and the narrow drop of the coherence coincide with the centroid frequency of the imaginary QPO. These frequencies are marked with vertical dashed red lines in Fig.~\ref{fig:lagycohe}. As we move further back to earlier observations, the amplitude of the QPO decreases in the imaginary part of the CS, while it increases in the real part. Eventually, this variability component becomes significant in the power spectra and is no longer `imaginary'. We are able to track the evolution of not just the QPO but also other Lorentzian components by fitting a consistent multi-Lorentzian model to all observations of the HSS-to-LHS transition (see Fig.~\ref{fig:allfits}). We note that these Lorentzian components jointly shift to higher frequencies for softer states of the source, preserving their relative order as they move in frequency. This consistency in the model allows us to identify the imaginary QPO as evolving from a `real' QPO observed in the PS during the earlier stages of the HSS-to-LHS transition. In Fig.\ref{fig:fit265}, we present the fit of the power spectra (left) and the real and imaginary parts of the CS (right) for observation 1200120265, with their corresponding residuals. In this case, the QPO, depicted by the vertical orange line, is needed to fit the PS. The derived models for the phase lags and coherence function (not shown) do not predict any narrow features, consistently with the observed behaviour in the earlier observations in Fig.~\ref{fig:lagycohe}. Henceforth, when describing the properties of the QPO across the earlier stages of HSS-to-LHS transition, where the imaginary part of the CS is smaller than the real one, we refer to it as the `real QPO'. 

Moving forwards in time from observation 1200120268, we again detect the imaginary QPO only in the CS at the frequency of the narrow features in the phase lags and coherence function. In particular, the characterisation of the imaginary QPO in observation 1200120270 is less certain due to the reduced number of 65-seconds segments available for the analysis (see Table \ref{tab:tabla}).

\subsection{Time-evolution of QPO properties}

On the left panels of Fig.~\ref{fig:rmsylagvst} we show the time-evolution of the rms amplitudes of the QPO in the hard (upper panel) and soft (bottom panel) energy bands, while on the right panel, we show the time-evolution of the phase lags of the QPO between those energy bands. The horizontal bars in the plot indicate the duration of each observation. In observation 1200120268, the rms amplitude of the imaginary QPO in the $2-12$~keV energy band is $\sim$3.4\%. As we move to earlier observations, the rms amplitude of the hard band increases up to $\sim$8.3\% for observation 1200120265. Instead, for even earlier observations, the rms decreases to $\sim$2.7\%. Regarding the $0.3-2$~keV energy band, the rms amplitude for observation 1200120268 is $\sim$1.4\%, and it remains more or less constant as we move back in time. For observation 1200120265, the rms amplitude rises to about $\sim$2.2\% and decreases to $\sim$0.7\% for earlier observations. On the right panel of Fig.~\ref{fig:rmsylagvst}, we observe that the phase lag for observation 1200120268 is $\sim$1.2~rad. We find smaller magnitudes of the phase lags corresponding to softer states of the source, consistent with the decreasing amplitude of the imaginary part and the increasing amplitude of the real part of the QPO. In observation 1200120264\_p1, the QPO has a soft phase lag of $\sim-0.4$~rad that, being less than $\pi/4$ in magnitude, indicates that its real part is larger than its imaginary part. On the other hand, for observation 1200120269, the imaginary QPO has a large hard lag of $\sim$1.9~rad, its rms amplitude in the hard energy band is $\sim$2.5\% and in the soft band, $\sim$1.1, respectively. For the following observation, the phase lag of the imaginary QPO decreases to $\sim$0.7~rad, while the rms amplitudes increase to $\sim$4\% and $\sim$2.1\%, respectively. We note that the centroid frequency of the QPO shows a general tendency to decrease as the source hardens, from $9.1$~Hz to $1.1$~Hz, as evidenced by the colour gradient in Fig.~\ref{fig:rmsylagvst}. A similar decreasing trend with frequency is observed for the other Lorentzians, which we can trace across the subsequent observations.

   \begin{figure}
   \centering
   \includegraphics[width=\columnwidth]{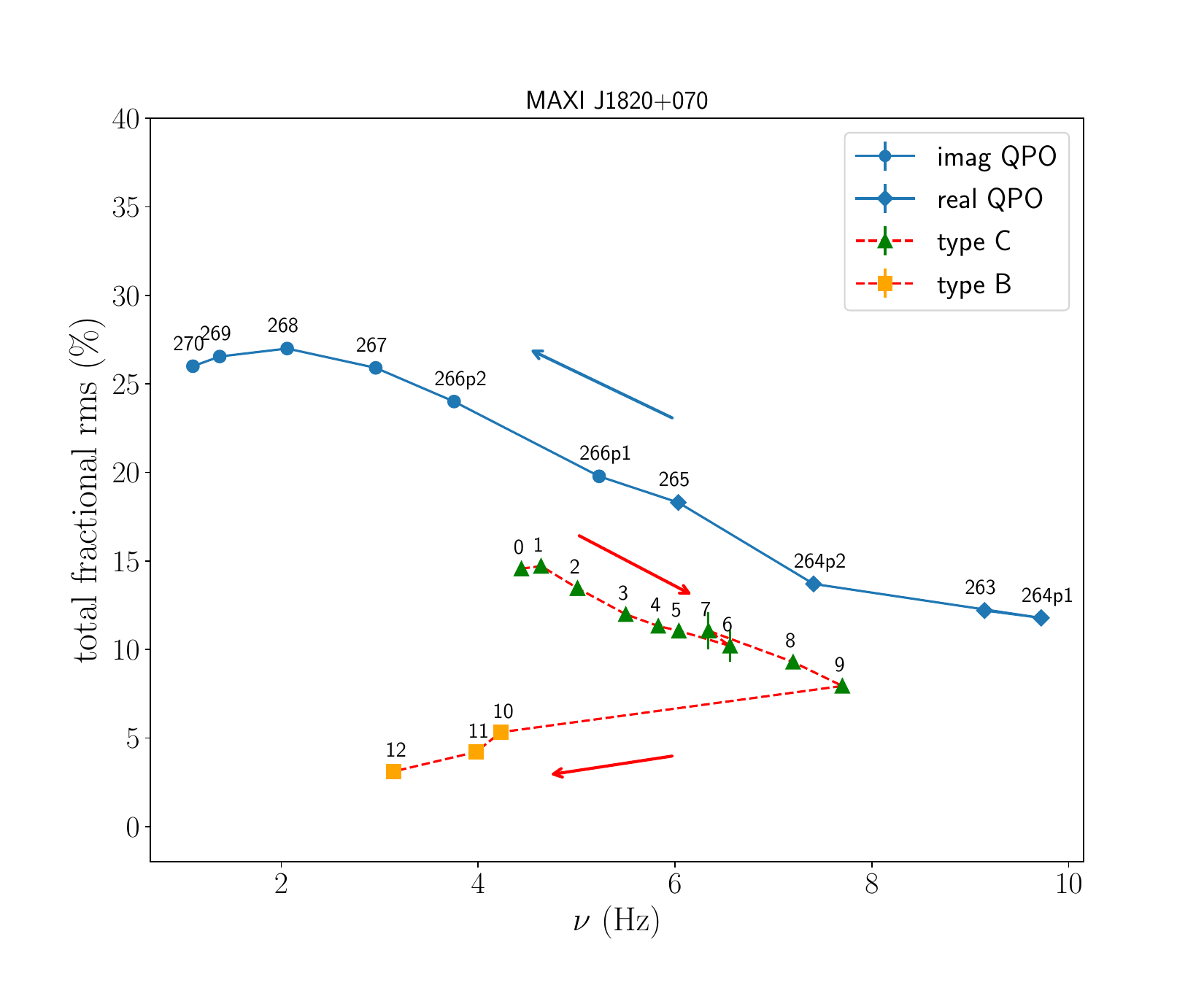}
      \caption{Total fractional rms in the 0.3-12~keV band integrated from 0.1~to~50~Hz versus centroid frequency of the QPO in \MAXI (blue), with both quantities derived from our fitted model. Diamonds correspond to observations 1200120263-1200120265 while dots correspond to observations 1200120266\_p1-1200120270. Green triangles and orange squares depict the type B and C QPOs identified in each orbit of observation 1200120197, which covered the HIMS-SIMS transition, studied by \citet{Ma2023}. The solid and dashed lines connect the QPOs during the source hardening and softening respectively. Arrows indicate the direction of time progression in each sequence.}
         \label{Fig:Motta}
   \end{figure}

\citet{Motta2011} found a relation between the amplitudes of the broadband variability of the source GX339--4 and the QPO frequency for different QPO types. In their Fig. 4, \citet{Motta2011} plotted the broadband fractional rms of the PS integrated between 0.1~Hz and 64~Hz vs. the QPO centroid frequency, using {\em RXTE} data in the $2-20$~keV energy range. They showed that different types of QPOs occupy different regions in the plot. In Fig.~\ref{Fig:Motta}, we reproduce that plot for the QPOs in \MAXI using \NICER data in the $0.3-12$~keV energy range and considering the $0.1-50$~Hz frequency range. The blue dots and diamonds correspond to the imaginary and the real QPOs, respectively, in our considered observations. We also include the type B (yellow squares) and C (green triangles) QPOs identified by \citet{Ma2023} in different orbits of observation 1200120197 of \MAXI. The arrows indicate the time progression of these measurements. The difference in the direction in which the source moves is because in the orbits of 1200120197 (red), \MAXI is softening while in our observations (blue) the source is hardening. We see that, as the source transitions towards the LHS, the QPO frequency decreases and the total variability increases from $\sim$11.7\% to $\sim$27\%. The rms amplitudes of the type C QPO span from $\sim$8\% to $\sim$15\% while those of the type B QPO are between $\sim$3\% and $\sim$5\%. For both type B and C QPOs we find smaller rms amplitudes than the values in GX\,339$-$4 in \citet{Motta2011}. 

In Table~\ref{tab:tabla} we give the best-fitting values of the real/imaginary QPO in \MAXI for each observation, along with their corresponding $1-\sigma$ uncertainties. For comparison, we also include the properties of typical type B and C QPOs \citep{Casella2005,Motta2016}. In \MAXI, as the source hardens, the centroid frequency of the real/imaginary QPO evolves from $\sim 9$ to $\sim 1$~Hz. The FWHM of the Lorentzian increases from $\sim$2~Hz to $\sim$3~Hz during the observations where no narrow features are seen in the coherence or phase lags, corresponding to a decreasing quality factor from $\sim$4.3 to $\sim$2. After observation 1200120266\_p1 the FWHM of this component decreases from $\sim$3 to $\sim$0.1~Hz. This leads to an increasing quality factor from $\sim$2 to $\sim$11.6. We also list the rms amplitudes of the imaginary QPO in the 2$-$12~keV band, which range from $\sim$2\% to $\sim$8\% as shown in Fig.~\ref{fig:rmsylagvst}, and the broadband rms amplitudes between $0.1-50$~Hz in the $0.3-12$~keV band that, as shown in Fig.~\ref{Fig:Motta}, increase from $\sim$12\% to $\sim$27\%.

\begin{table*}
\centering
 \caption{Best-fitting values of the QPO in \MAXI for each observation with their corresponding 1-$\sigma$ uncertainties.}
 \begin{tabular}{cccccccc}
  \hline
  \noalign{\smallskip}
  Observation & Start time & $\nu_{\rm QPO}$ & FWHM & $Q= (\nu_0/\mathrm{FWHM})$ & ${\rm rms_{~QPO}^a}$ & ${\rm rms_{0.1-50~Hz}^b}$ & 65-sec segments\\
  & MJD$-$58385 & (Hz) & (Hz) & & (\%) & (\%) &   \\
  \noalign{\smallskip}
  \hline
  \noalign{\smallskip}
  1200120263\phantom{\_p1} & 0.620 & 9.14~$\pm$~0.29 & 2.1~$\pm$~0.8 & 4.3~$\pm$~1.8 & 2.7~$\pm$~0.5 & 12.2~$\pm$~0.2 & 12\\
  1200120264\_p1 & 1.196 & 9.72~$\pm$~0.25 & 2.8~$\pm$~0.9 & 3.5~$\pm$~1.2 & 2.7~$\pm$~0.4 & 11.8~$\pm$~0.1 & 16\\
  1200120264\_p2 & 1.840 & 7.41~$\pm$~0.14 & 3.0~$\pm$~0.5 & 2.4~$\pm$~0.4 & 6.3~$\pm$~0.5 & 13.7~$\pm$~0.3 & 6\\
  1200120265\phantom{\_p1} & 2.417 & 6.03~$\pm$~0.08 & 3.1~$\pm$~0.3 & 1.9~$\pm$~0.2 & 8.3~$\pm$~0.4 & 18.3~$\pm$~0.2 & 11\\
  1200120266\_p1 & 3.191 & 5.23~$\pm$~0.05 & 1.2~$\pm$~0.2 & 4.3~$\pm$~0.8 & 4.2~$\pm$~0.7 & 20.0~$\pm$~0.2 & 17 \\
  1200120266\_p2 & 3.516 & 3.75~$\pm$~0.03 & 0.5~$\pm$~0.1 & 7.3~$\pm$~1.5 & 3.9~$\pm$~0.8 & 24.4~$\pm$~0.2 & 27 \\
  1200120267\phantom{\_p1} & 4.156 & 2.96~$\pm$~0.06 & 0.9~$\pm$~0.2 & 3.4~$\pm$~1.0 & 3.4~$\pm$~1.1 & 26.5~$\pm$~0.2 & 40 \\
  1200120268\phantom{\_p1} & 4.993 & 2.06~$\pm$~0.03 & 0.3~$\pm$~0.1 & 6.3~$\pm$~2.8 & 2.4~$\pm$~0.8 & 27.4~$\pm$~0.2 & 54\\
  1200120269\phantom{\_p1} & 6.216 & 1.37~$\pm$~0.04 & 0.21~$\pm$~0.2 & 6.3~$\pm$~4.8 & 2.3~$\pm$~2.0 & 27.0~$\pm$~0.8 & 39\\  
  1200120270\phantom{\_p1} & 7.053 & 1.10~$\pm$~0.01 & 0.1~$\pm$~0.1 & 11.6~$\pm$~5.0 & 3.9~$\pm$~2.0 & 26.0~$\pm$~0.3 & 14\\
  \noalign{\smallskip}
  \hline
  \noalign{\smallskip}
  Type B & & $\sim5-6$ & & $\gtrsim6$ & \\  
  \noalign{\smallskip}
  \hline
  \noalign{\smallskip}
  Type C & & $\sim0.1-15$ & & $\sim7-12$& \\
  \noalign{\smallskip}
  \hline
 \end{tabular}
 \flushleft{$^a$ The rms amplitude of the QPO in the $2.0-12.0$~keV band.\\
 $^b$ The broadband ($0.1-50$~Hz) rms amplitude in the $0.3-12.0$~keV band.}
 \tablefoot{For comparison, we also include the properties of typical type B and C QPOs \citep{Casella2005,Motta2016}, but see \citet{Alabarta2022} and \citet{Ma2023} where they found type-C QPOs with Q factors between 0.2 and 7 in MAXI~J1348--630 and \MAXI, respectively. In the last column, we provide the number of 65-second segments available for the analysis of each observation.}
 \label{tab:tabla}
\end{table*}

\subsection{Energy dependence of rms and phase lags of the QPO}

\begin{figure*}
   \centering
   \includegraphics[width=0.9\textwidth]{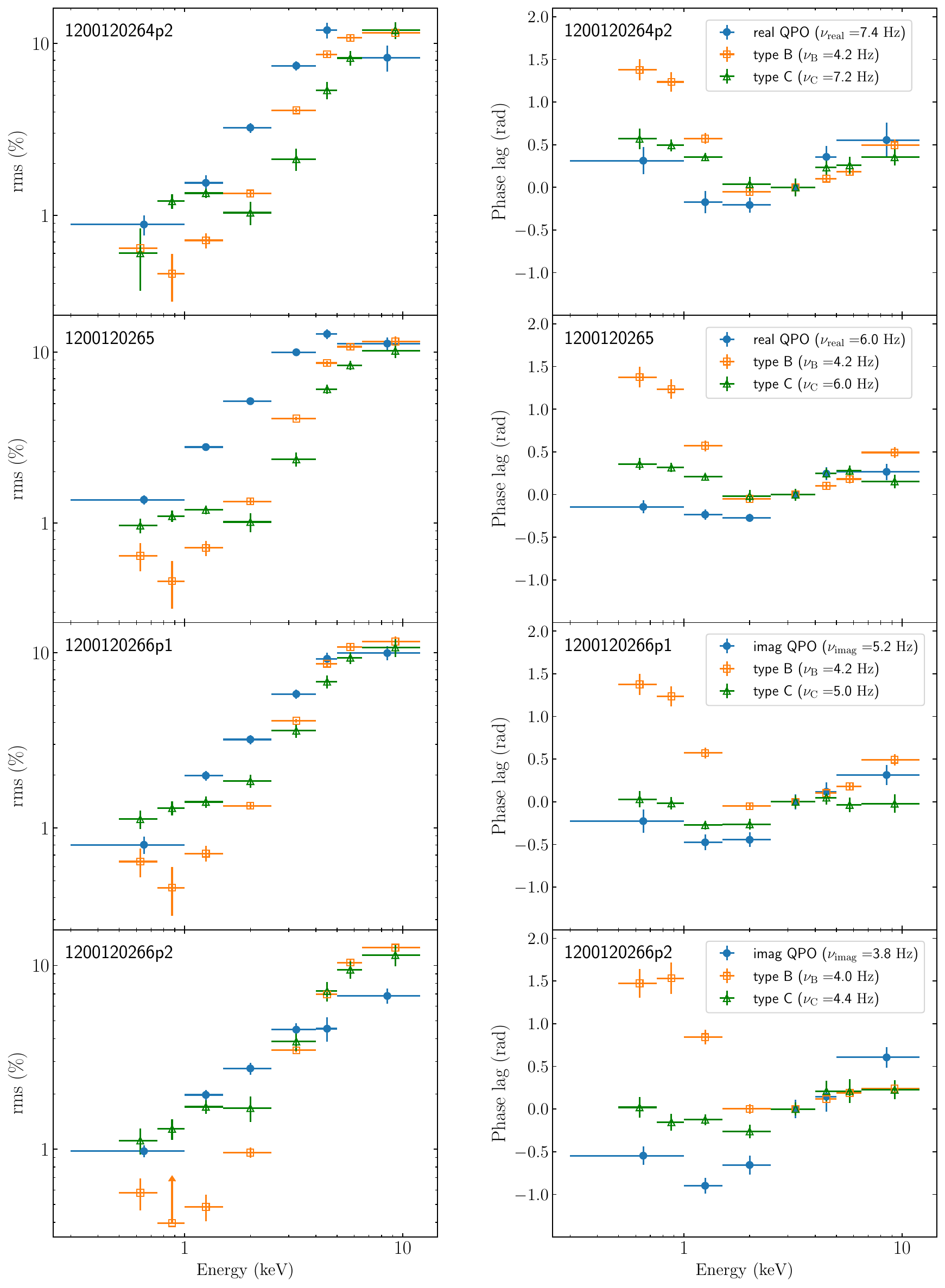}
   \caption{The rms (left panels) and phase-lag (right panels) spectra of the imaginary/real (blue circles), type-B (orange squares) and type-C (green triangles) QPOs in \MAXI. From top to bottom, the panels correspond to observations 1200120264\_p2, 1200120265, 1200120266\_p1 and 1200120266\_p2 of \MAXI. The imaginary/real QPOs are from this work, while the type B and C QPOs are from observation 1200120197 analysed by \cite{Ma2023}. For the comparison, we selected type-B and C QPOs with centroid frequencies most similar to those of the imaginary/real QPOs. Horizontal lines cover the range of the energy band corresponding to each marker.}
              \label{fig:rmsandlagspectra}%
    \end{figure*}

Finally, to explore the energy dependence of the rms and phase lags of the imaginary QPO, we simultaneously fit the PS in each of the energy bands mentioned in Sec.~\ref{sec:obs} with the PS in the total band, $0.3-12$~keV, and the CS between them. For this, we use the best-fitting model obtained for the corresponding observation and we fix the centroid frequencies and FWHM of every Lorentzian. In Fig.~\ref{fig:rmsandlagspectra}, we show the rms (left) and phase-lag (right) spectra of the real/imaginary QPO for observations 1200120264\_p2, 1200120265, 1200120266\_p1, and 1200120266\_p2 (see Fig.~\ref{fig:allrmsylags} for the spectra of all the observations of the HSS-to-LHS transition). We depict these spectra with blue dots and we also include the spectra of a type-B (orange squares) and a type-C (green triangles) QPO from \citet{Ma2023}.  We choose these QPOs for comparison because their centroid frequencies are the closest to that of the imaginary QPO in each observation. 

On the one hand, we note that the rms spectra in Fig.~\ref{fig:rmsandlagspectra} generally exhibit similar patterns across cases, always increasing with energy. The rms amplitude of the type-B QPOs remains relatively constant at $\sim$0.6\% between 0.5 and 2~keV, then rises to $\sim$10\%. The rms of the real/imaginary QPO and the type-C QPOs increase with energy from $\sim$1\% at $\sim$0.5 keV to $\sim$10-13\% at $\sim$10 keV. However, in the case of observations 1200120264\_p2 and 1200120265, the rms of the real QPO reaches higher values compared to the type-B and type-C QPOs, except for the last energy bin where the rms of the real QPO drops from $\sim12\%$ to $\sim8\%$ for observation 1200120264\_p2, and from $\sim12.7\%$ to $\sim11.2\%$ for observation 1200120265 For observation 1200120266\_p2 at energies above $\sim$3~keV the rms of the imaginary QPO remains more or less constant around $\sim$5\%. 

On the other hand, we see significant differences in the phase-lag spectra in Fig.~\ref{fig:rmsandlagspectra}. We shift the phase lags such that the subject band $2.5-4$~keV works as the reference band in the plot. We note that it is the shape of the lag spectrum that matters rather than the specific values of the lags, since these are relative quantities. We observe that, at higher energies, the behaviour is more or less consistent with a hardening trend, except for the type-C QPO with centroid frequency of 5~Hz, used for comparison in the third panel, whose lags remain more or less constant. At lower energies, the lags of type-B QPOs increase to $\sim$1.5~rad as energy decreases, while for type-C QPOs, the lags increase less, up to 0.5~rad, or remain relatively constant around $\sim$0~rad \citep{Ma2023}. In the four panels of Fig.~\ref{fig:rmsandlagspectra}, we see sort of U-shaped patterns across the full energy range for the phase-lag spectra of the real/imaginary QPO. For observation 120012064\_p2, this U-shape spans a total range of $\sim$0.7~rad, it has a minimum lag of $\sim-0.2$~rad at the third energy channel ($1.5-2.5$~keV) and it is very similar to the phase-lag spectra of the type-C QPO. For observation 1200120265, the lag spectrum of the real QPO has a minimum of $\sim-0.3$~rad at the third energy channel, and for lower energies, the lags slightly increase up to $\sim-0.1$~rad. In the third and fourth panels, we see that the U-shaped patterns for the imaginary QPOs span a total range of $\sim$0.8~rad for observation 1200120266\_p1, from $\sim -0.5$~rad to $\sim$0.3~rad, and of $\sim$1.4~rad for observation 1200120266\_p2, from $\sim -0.9$~rad to $\sim$0.5~rad.

\section{Discussion}
\label{sec:discussion}
We discovered imaginary quasi-periodic oscillations (QPOs) in NICER observations during the high-soft to low-hard states transition of the black hole candidate MAXI J1820+070 by fitting simultaneously power spectra (PS) and the cross-spectrum (CS) of the source. The imaginary QPOs appear as narrow features with a small amplitude in the Real and a large one in the Imaginary parts of the CS; these QPOs are not significantly detected in the PS if they overlap in frequency with other variability components (see Sec.~\ref{sec:math}). The coherence function drops and the phase lags increase abruptly at the frequency of the imaginary QPO (see Figs.~\ref{fig:lagycohe} and \ref{fig:fit268}). Furthermore, we detect a QPO that is significant in the power spectra in earlier observations in the outburst after the source left the high-soft state. This QPO evolves into the imaginary QPO as the source hardens. We compare the properties of the imaginary QPO with those of the typical type B and C QPOs previously observed in the rise of the outburst of this source and we find similarities with type-C QPOs, while we cannot rule out that the imaginary one is a new type of QPO.

We confirm the assessment of \cite{Mendez2024}, that by looking for variability components only in the PS, as has so far been done, we can miss significant signals that have a large Imaginary and a small Real part of the CS. This type of signal can be overshadowed by other components in the PS, but is significantly detected by analysing the CS, where they have a higher SNR (see Eq.~\ref{eq:SNRcs} and \ref{eq:SNRps}). On the other hand, the coherence function, when close to unity, would be more sensitive than both the PS and CS, which emphasises the relevance of using the coherence function to identify variability components. 

During outburst, \MAXI traced an overall \textit{q} shaped path in the HID, from the hard state to the soft state and back to the hard state \citep{Shidatsu2019}, with notable transitions in count rates and spectral hardness ratios. We examine the phase-lag and coherence function between photons in the $0.3-2$~keV (soft) and $2-12$~keV (hard) bands across all 131 observations. Among these, we identify five observations during the decay of the outburst (obsID $1200120266-1200120270$) in which a significant drop in the coherence coincided with a feature in the phase-lag spectrum where, as the frequency increases, the lags increase sharply and then decrease smoothly \citep[firstly observed by][in obsID 1200120268]{Konig2024}. We call this feature the cliff, because of its shape. Applying the technique introduced by \cite{Mendez2024}, we find that the drop in the coherence function and the cliff in the phase-lag spectrum occurred at the frequency at which we needed to add a narrow Lorentzian to fit the shape of the Imaginary part of the CS. The frequency of this Lorentzian, which represents the imaginary QPO, decreases as the source hardens. \citet{Konig2024} analysed NICER data from Cyg~X-1 and found that the source occupies the lower branch of the $q$ diagram during those observations (and also all the observations with the {\em Rossi X-ray Timing Esplorer}), and that the coherence function exhibited a narrow drop that coincided with an abrupt increase in the lags. The authors also noticed that softer states of the source corresponded to higher values of the frequency of this timing feature (see their Fig. 15). \MAXI is the second source to show the same phenomenology. As in Cyg~X-1, the sharp increase of the lags and the narrow drop of the coherence happen in the lower branch of the $q$ in the HID, and the frequency of the feature decreases as the source spectrum hardens. \cite{Konig2024} also observed these narrow features in one observation of the BHXB MAXI~J1348-630 (see their Fig.~17). In their Fig.~7, \cite{Alabarta2024} found the drop in the coherence and the sudden increase in the phase lags coinciding with a type-C QPO of decreasing frequency in three observations during the decay of the MAXI~J1348--630 outburst \citep[one of which was the same observation presented in][]{Konig2024}.

We also include observations 1200120263-1200120265 in our analysis, when the source just left the HSS and began its transition towards the LHS. Applying the same technique to fit those data, we find a component that resembles the imaginary QPO but is also significant in the PS. At those observations, the QPO has a large real part and a small imaginary part, akin to the other variability components, which corresponds to a phase lag of $\sim$0~rad at those frequencies, consistent with the absence of the cliff in the corresponding panels of Fig.~\ref{fig:lagycohe}. Furthermore, the QPO dominates the variability at these frequencies, so the coherence function remains near unity due to the lack of significant overlap with other components. Using the model initially fitted to observation 1200120268, where the imaginary QPO is insignificant in the PS, we tracked this QPO backwards in time. As we analysed earlier observations, corresponding to softer states of the source, we identified the same Lorentzian components gradually shifting to higher frequencies. This allowed us to observe the evolution of the QPO from being significant in the PS to becoming less significant as its imaginary part grows (as we showed in Sec.~\ref{sec:math}), eventually producing the narrow features in the phase lags and coherence as the source hardens.

\citet{Kawamura2023} proposed a model of propagating fluctuations \citep{Lyubarskii1997,Arevalo2006} to reproduce the variability in \MAXI, assuming that QPOs are multiplicative modulations of the spectral components included in the model. Their proposal stemmed from the fact that they found no significant features in the phase-lag spectra at the QPO frequency. The cliff that we identified in the phase-lag spectra and the simultaneous drop in the coherence function, however, challenge their assumption. Indeed, the presence of these features indicates that QPOs cannot be solely described as a multiplicative process within the hot flow, but instead points out to an additional variability component affecting the phase-lags and coherence function. This implies that the scenario by \citet{Kawamura2023} has to incorporate additive components to model QPOs and fully account for the observed timing properties of \MAXI. 

The drop in the coherence function and the cliff in the phase-lag spectrum are not significantly detected when we consider the PS and CS between two hard energy bands, $2-5$~keV and $5-12$~keV \citep[the same was seen for the Cyg X-1 data by][see their Fig. 10]{Konig2024}. Correspondingly, the imaginary QPO is not significant in either the PS or the CS between those hard energy bands.  Since a soft energy band is needed to have those features, the imaginary QPO may correspond to a variable component at low energies and, therefore, it may be linked to the accretion disc. Furthermore, the imaginary QPO appears in the lower branch of the {\it q} in the HID \citep{Konig2024}, when the source has left the HSS and transitions towards the LHS. This could be due to the expansion of the truncated accretion disc inner radius or to the reappearance of the corona extending over the disc \citep{Peng2023} and potentially leading to feedback \citep{Karpouzas2020,Bellavita2022}.

\citet{Veledina2016} proposed that the peaked noise in the PS of BHXBs could be the result of the interference of disc Comptonization and synchrotron Comptonization, both modulated by accretion rate fluctuations and separated by a time delay. We explore whether the peaked broadband variability in the PS of \MAXI could arise from the interference of two broad variability components. Such a fit, however, leads to a wide and shallow drop in the coherence function \citep{Nowak1999,NowakNWD,Nowak2000}, rather than the deep and narrow drop observed in the data. To recover the sharp feature in the coherence function, it is necessary to add the narrow imaginary QPO. In their Fig.~16, \citet{Konig2024} showed the fitting results applying \citet{Mendez2024} technique and they also noticed that, in order to reproduce the abrupt increase of the lags and the narrow drop in the coherence observed in the data, they needed to add a narrow component at the frequency of the timing feature. These findings challenge the interference idea since that mechanism alone cannot explain the results of \cite{Konig2024} for Cyg~X-1, and our results for \MAXI.

The need to include a narrow Lorentzian evidences that the corresponding imaginary QPO is an independent component of variability in the PS and the CS \citep{Mendez2024}. The model used to predict the phase lag and coherence function is based on the assumption that the variability components are mutually incoherent. The accurate prediction of these frequency spectra, though not a proof of this hypothesis, adds confidence to the underlying assumption. Therefore, we suggest that the other Lorentzian components, besides the imaginary QPO, may be independent of each other, and each of them may arise from different independent processes \citep[e.g., ][]{Nowak1997,Mendez2013}. Whether this is the case for the other Lorentzians or just for the imaginary QPO, we conclude that the abrupt cliff in the phase lags and the narrow drop in the coherence cannot be described with a single monolithic function \citep{Reynolds1999}. This means that each variability component, which appears to represent an individual physical process, requires its own transfer function and therefore the variability cannot be explained by a single, global transfer function. 

We study the time-evolution of the rms amplitudes, both in the hard and soft energy bands, and of the phase lags between the same two energy bands, of the QPO. During the first stages of the HSS-to-LHS transition, when the QPO is also significant in the PS, the rms of the QPO in the 2$-$12~keV band increases until it reaches $\sim$8\%. At these stages of the transition, the rms amplitude in the 0.3$-$2~keV band increases from $\sim$0.7\% to $\sim$2.2\%. Then the rms amplitudes of the imaginary QPO decrease to $\sim$2\% in the hard band and $\sim$1.3\% in the soft band. On the other hand, the phase lags increase from $\sim-0.4$~rad to $\sim$1.9~rad as the source spectrum hardens. In particular, for observations from 1200120263 to 1200120265 the phase lags of the QPO are between $-0.4$ and 0.2~rad  and therefore less than $\pi/4$ in magnitude, which is consistent with the fact that this component is not hidden in the PS, since its real part in the CS is larger than its imaginary part.

We plot the broadband fractional rms in the PS in the $0.3-12$~keV energy band integrated from 0.1~Hz to 50~Hz vs. the QPO frequency, emulating the plot in Fig.~4 of \citet{Motta2011}. We also include the type B and C QPOs identified by \citet{Ma2023} in different orbits of observation 1200120197 of \MAXI. The rms values for the imaginary QPO are larger than those of both types of QPOs in observation 1200120197. However, it was observed in different sources that the hardening phase of an outburst shows larger rms than the softening phase \citep{Munoz2011,Motta2011}. The broadband rms is influenced by the multiple variability components in the PS, which results in the rms reflecting the overall variability integrated across the wide frequency range, rather than the contribution from only a single QPO. Consequently, while the broadband rms provides useful information about the general variability state of the source, it does not allow for a clear distinction between different types of QPOs. Therefore, it is not feasible to attribute a QPO type based solely on this plot.

We compare the best-fitting values of the centroid frequency and the $Q$ factor of the QPO for each observation to those of the typical type B and C QPOs. While the frequency range of the QPO during the HSS-to-LHS transition is consistent with a type C QPO, its $Q$ factor is smaller than the expected range for the typical QPO types. However, using NICER observations, type-C QPOs with $Q$ factors between 0.2 and 9 have been detected during the rise of the outburst of \MAXI by \citet{Ma2023} and during the decay of the outburst and the reflare of the BHXB MAXI~J1348--630 by \citet{Alabarta2022}.

Finally, we analyse the energy-dependent rms and phase lags spectra of the QPO and compare them to those of a type B and a type C in \MAXI studied by \cite{Ma2023}. We find that the rms spectra of all QPOs exhibit the same general tendency to increase with energy, but below $1.5-2$~keV the imaginary and the type-C QPOs show a larger rms amplitude than the type B QPO. For observations 1200120264\_p2 and 1200120265, we observe higher values for the rms of the imaginary QPO than for those of the type-C QPOs used for comparison, except for the energy channel $5-12$~keV where the rms amplitude of the imaginary QPOdrops from $\sim12\%$ to $\sim8\%$ and from $\sim12.7\%$ to $\sim11.2\%$, respectively. We note that the phase lag spectra, above the reference energy band, $2.5-4$~keV, have similar patterns consistent with hardening trends, except for the type-C QPO used for comparison with the imaginary QPO in 1200120266\_p1, whose phase lags remain more or less constant. However, below 2~keV, the phase lags diverge: the lags of type B and C QPOs increase as energy decreases, with type B QPO showing the largest magnitude of the lags, while the lags of the imaginary QPO decrease for lower energies. The phase-lag spectra of the real and imaginary QPOs describe a U-shape pattern across the full energy range. In particular, in observations 1200120264\_p2 and 1200120265, though slightly harder, the U-shapes of the real QPO are very similar to those of type-C QPOs in the hard-to-soft transition, at the rising part of the outburst. In these two observations, we note that averaging the lags below 2~keV and those above 2~keV, the phase-lag difference between them is $\sim0$~rad, and we recover the fact that the real QPO has a small imaginary part of the CS. In observation 1200120266\_p1, the U-shape pattern of the imaginary QPO resembles that of the type-C QPO below 2~keV. However, at higher energies, the two differ: the lags of the imaginary QPO show a hardening trend while those of the type-C QPO remain more or less constant. The phase-lag spectrum of the imaginary QPO in observation 1200120266\_p2 strongly hardens for energies below 2~keV, diverging from that of the type-C QPO, and the U-shaped pattern described covers a total range of $\sim$1.5~rad, from $\sim -0.9$~rad to $\sim$0.5~rad. The phase difference between the average lags below and above 2~keV is $\sim0.3$~rad for observation 1200120266\_p1 and $\sim0.6$~rad for observation 1200120266\_p2, thus, we recover the fact that the imaginary QPO has a significant imaginary part in the CS. A similar U shape to that of the imaginary QPO was observed for a QPO of $Q\sim2$ classified as type-C during the decay of the outburst of MAXI~J1348--630 \citep[see panel (b) of Fig.~10 in][]{Alabarta2022}, spanning a significant range of $\sim$0.7~rad in the phase-lags spectrum. In their Fig.~7, \cite{Alabarta2024} showed that, at the frequency of this type-C QPO, the coherence drops and the phase lags suddenly increase \citep[these features in MAXI~J1348--630 were firstly observed by][in the third column of their Fig.~17]{Konig2024}. This suggests that the imaginary QPO might indeed be a type-C QPO. If this were the case, then given that type-C QPOs are found in the HIMS and LHS, detecting the real/imaginary QPO just six days after the source left the HSS would suggest either an extremely short SIMS or its complete absence after the HSS. 

In summary, during the hardening phase of \MAXI outburst, we identify an imaginary QPO whose properties resemble those of observed type-C QPOs. This imaginary QPO evolves from a real QPO, likely a type-C QPO, as the source transitions towards harder states. As this evolution occurs, the real part of the CS of this Real QPO decreases whereas its Imaginary part increases, and eventually it becomes the imaginary QPO. Due to the overlapping in frequency with other variability components and the signal-to-noise ratio, the imaginary QPO ends up hidden in the PS. Thus, the imaginary QPO is not detected in the PS, but it is significant in the CS and it is unveiled by the narrow features that occur at the QPO frequency: the cliff in the phase lags and the narrow drop in the coherence function. Given that the imaginary QPO is only observed when considering a soft energy band below $2$~keV, it may correspond to a component variable at low energies, likely linked to the accretion disc. This QPO is detected when the source is in an intermediate state, transitioning from the HSS towards the LHS, suggesting it could be related to the expansion of the inner radius of the truncated disc or to the feedback due to the expansion of the corona over the disc. As the source transitions towards the LHS, the frequency of the imaginary QPO decreases, and so does the rms amplitude of the broadband variability in the $0.3-12$~keV band.  The centroid frequency of this QPO is consistent with those of typical type-C QPOs, while the $Q$ factors are smaller than the typical ones. However, using \NICER observations, \citet{Alabarta2022} and \citet{Ma2023} observed type-C QPOs with similar $Q$ values in MAXI~J1348--630 and \MAXI, respectively. The energy-dependent rms spectra of the imaginary QPO exhibit similar trends to those of the type-C QPO observed in \MAXI by \citet{Ma2023}. For both the imaginary and type-C QPOs, the phase lags are hard at energies above $\sim$2~keV. However, at lower energies, the phase lags of the imaginary QPO are harder. To better understand the imaginary QPO, further analysis of other sources during the decay of the outburst is necessary, particularly considering soft X-ray data. Such analysis is possible due to the large effective area and the timing capabilities of \NICER below 2~keV. While the imaginary QPO might represent a new type of QPO, it is also plausible that it is a type-C QPO, with the observed differences in its properties likely due to the position of the source in the HID, as it transitions from the HSS towards the LHS.

\begin{acknowledgements}
We thank the referee for the insightful suggestions that have improved the clarity of our work. CB thanks Federico Fogantini for the script of Fig.~\ref{fig:fit268} and for his valuable help with the plots. CB is a Fellow of CONICET. FG is a CONICET researcher. FG acknowledges support by PIBAA 1275 (CONICET). FG was also supported by grant PID2022-136828NB-C42 funded by the Spanish MCIN/AEI/ 10.13039/501100011033 and “ERDF A way of making Europe”. CB and FG acknowledge support by PIP 0113 (CONICET). CB, FG and MM acknowledge the research programme Athena with project number 184.034.002, which is (partly) financed by the Dutch Research Council (NWO). RM acknowledges support from the China Postdoctoral Science Foundation (2024M751755). OK acknowledges NICER GO funding 80NSSC23K1660. 
\end{acknowledgements}




\bibliographystyle{aa}
\bibliography{aanda} 

\begin{thebibliography}{65}
\expandafter\ifx\csname natexlab\endcsname\relax\def\natexlab#1{#1}\fi

\bibitem[{{Adachi} {et~al.}(2020){Adachi}, {Murata}, {Oeda}, {Niwano},
  {Shiraishi}, {Iida}, {Ogawa}, {Hosokawa}, {Nakamura}, {Toma}, {Yatsu}, \&
  {Kawai}}]{Adachi2020}
{Adachi}, R., {Murata}, K.~L., {Oeda}, M., {et~al.} 2020, The Astronomer's
  Telegram, 13502, 1

\bibitem[{Alabarta {et~al.}(2024)Alabarta, M{\'e}ndez, Garcia, Altamirano,
  Zhang, Zhang, Russell, \& K{\"o}nig}]{Alabarta2024}
Alabarta, K., M{\'e}ndez, M., Garcia, F., {et~al.} 2024, Astrophysical Journal,
  arXiv:2409.14883

\bibitem[{{Alabarta} {et~al.}(2022){Alabarta}, {M{\'e}ndez}, {Garc{\'\i}a},
  {Peirano}, {Altamirano}, {Zhang}, \& {Karpouzas}}]{Alabarta2022}
{Alabarta}, K., {M{\'e}ndez}, M., {Garc{\'\i}a}, F., {et~al.} 2022, \mnras,
  514, 2839

\bibitem[{{Ar{\'e}valo} \& {Uttley}(2006)}]{Arevalo2006}
{Ar{\'e}valo}, P. \& {Uttley}, P. 2006, \mnras, 367, 801

\bibitem[{{Bahramian} \& {Degenaar}(2023)}]{Bahramian2023}
{Bahramian}, A. \& {Degenaar}, N. 2023, in Handbook of X-ray and Gamma-ray
  Astrophysics. Edited by Cosimo Bambi and Andrea Santangelo, 120

\bibitem[{{Bellavita} {et~al.}(2022){Bellavita}, {Garc{\'\i}a}, {M{\'e}ndez},
  \& {Karpouzas}}]{Bellavita2022}
{Bellavita}, C., {Garc{\'\i}a}, F., {M{\'e}ndez}, M., \& {Karpouzas}, K. 2022,
  \mnras, 515, 2099

\bibitem[{{Belloni} {et~al.}(2005){Belloni}, {Homan}, {Casella}, {van der
  Klis}, {Nespoli}, {Lewin}, {Miller}, \& {M{\'e}ndez}}]{Belloni2005}
{Belloni}, T., {Homan}, J., {Casella}, P., {et~al.} 2005, \aap, 440, 207

\bibitem[{{Belloni} {et~al.}(2002){Belloni}, {Psaltis}, \& {van der
  Klis}}]{Belloni2002}
{Belloni}, T., {Psaltis}, D., \& {van der Klis}, M. 2002, \apj, 572, 392

\bibitem[{{Belloni}(2010)}]{Belloni2010}
{Belloni}, T.~M. 2010, in Lecture Notes in Physics, Berlin Springer Verlag, ed.
  T.~{Belloni}, Vol. 794, 53

\bibitem[{{Belloni} {et~al.}(2024){Belloni}, {M{\'e}ndez}, {Garc{\'\i}a}, \&
  {Bhattacharya}}]{Belloni2024}
{Belloni}, T.~M., {M{\'e}ndez}, M., {Garc{\'\i}a}, F., \& {Bhattacharya}, D.
  2024, \mnras, 527, 7136

\bibitem[{{Belloni} \& {Motta}(2016)}]{Belloni2016}
{Belloni}, T.~M. \& {Motta}, S.~E. 2016, in Astrophysics and Space Science
  Library, Vol. 440, Astrophysics of Black Holes: From Fundamental Aspects to
  Latest Developments, ed. C.~{Bambi}, 61

\bibitem[{{Belloni} {et~al.}(2011){Belloni}, {Motta}, \&
  {Mu{\~n}oz-Darias}}]{Belloni2011}
{Belloni}, T.~M., {Motta}, S.~E., \& {Mu{\~n}oz-Darias}, T. 2011, Bulletin of
  the Astronomical Society of India, 39, 409

\bibitem[{{Belloni} \& {Stella}(2014)}]{Belloni2014}
{Belloni}, T.~M. \& {Stella}, L. 2014, \ssr, 183, 43

\bibitem[{Bendat \& Piersol(2011)}]{bendat2011random}
Bendat, J. \& Piersol, A. 2011, Random Data: Analysis and Measurement
  Procedures, Wiley Series in Probability and Statistics (Wiley)

\bibitem[{{Casella} {et~al.}(2004){Casella}, {Belloni}, {Homan}, \&
  {Stella}}]{Casella2004}
{Casella}, P., {Belloni}, T., {Homan}, J., \& {Stella}, L. 2004, \aap, 426, 587

\bibitem[{{Casella} {et~al.}(2005){Casella}, {Belloni}, \&
  {Stella}}]{Casella2005}
{Casella}, P., {Belloni}, T., \& {Stella}, L. 2005, \apj, 629, 403

\bibitem[{{Fender} {et~al.}(2004){Fender}, {Belloni}, \& {Gallo}}]{Fender2004}
{Fender}, R.~P., {Belloni}, T.~M., \& {Gallo}, E. 2004, \mnras, 355, 1105

\bibitem[{{Gendreau} {et~al.}(2016){Gendreau}, {Arzoumanian}, {Adkins},
  {Albert}, {Anders}, {Aylward}, {Baker}, {Balsamo}, {Bamford}, {Benegalrao},
  {Berry}, {Bhalwani}, {Black}, {Blaurock}, {Bronke}, {Brown}, {Budinoff},
  {Cantwell}, {Cazeau}, {Chen}, {Clement}, {Colangelo}, {Coleman},
  {Coopersmith}, {Dehaven}, {Doty}, {Egan}, {Enoto}, {Fan}, {Ferro}, {Foster},
  {Galassi}, {Gallo}, {Green}, {Grosh}, {Ha}, {Hasouneh}, {Heefner}, {Hestnes},
  {Hoge}, {Jacobs}, {J{\o}rgensen}, {Kaiser}, {Kellogg}, {Kenyon}, {Koenecke},
  {Kozon}, {LaMarr}, {Lambertson}, {Larson}, {Lentine}, {Lewis}, {Lilly},
  {Liu}, {Malonis}, {Manthripragada}, {Markwardt}, {Matonak}, {Mcginnis},
  {Miller}, {Mitchell}, {Mitchell}, {Mohammed}, {Monroe}, {Montt de Garcia},
  {Mul{\'e}}, {Nagao}, {Ngo}, {Norris}, {Norwood}, {Novotka}, {Okajima},
  {Olsen}, {Onyeachu}, {Orosco}, {Peterson}, {Pevear}, {Pham}, {Pollard},
  {Pope}, {Powers}, {Powers}, {Price}, {Prigozhin}, {Ramirez}, {Reid},
  {Remillard}, {Rogstad}, {Rosecrans}, {Rowe}, {Sager}, {Sanders}, {Savadkin},
  {Saylor}, {Schaeffer}, {Schweiss}, {Semper}, {Serlemitsos}, {Shackelford},
  {Soong}, {Struebel}, {Vezie}, {Villasenor}, {Winternitz}, {Wofford},
  {Wright}, {Yang}, \& {Yu}}]{Gendreau2016}
{Gendreau}, K.~C., {Arzoumanian}, Z., {Adkins}, P.~W., {et~al.} 2016, in
  Society of Photo-Optical Instrumentation Engineers (SPIE) Conference Series,
  Vol. 9905, Space Telescopes and Instrumentation 2016: Ultraviolet to Gamma
  Ray, ed. J.-W.~A. {den Herder}, T.~{Takahashi}, \& M.~{Bautz}, 99051H

\bibitem[{{Hambsch} {et~al.}(2019){Hambsch}, {Ulowetz}, {Vanmunster}, {Cejudo},
  \& {Patterson}}]{Hambsch2019}
{Hambsch}, J., {Ulowetz}, J., {Vanmunster}, T., {Cejudo}, D., \& {Patterson},
  J. 2019, The Astronomer's Telegram, 13014, 1

\bibitem[{{Homan} \& {Belloni}(2005)}]{Homan2005}
{Homan}, J. \& {Belloni}, T. 2005, \apss, 300, 107

\bibitem[{{Homan} {et~al.}(2020){Homan}, {Bright}, {Motta}, {Altamirano},
  {Arzoumanian}, {Basak}, {Belloni}, {Cackett}, {Fender}, {Gendreau}, {Kara},
  {Pasham}, {Remillard}, {Steiner}, {Stevens}, \& {Uttley}}]{Homan2020}
{Homan}, J., {Bright}, J., {Motta}, S.~E., {et~al.} 2020, \apjl, 891, L29

\bibitem[{{Homan} {et~al.}(2001){Homan}, {Wijnands}, {van der Klis}, {Belloni},
  {van Paradijs}, {Klein-Wolt}, {Fender}, \& {M{\'e}ndez}}]{Homan2001}
{Homan}, J., {Wijnands}, R., {van der Klis}, M., {et~al.} 2001, \apjs, 132, 377

\bibitem[{{Ingram}(2019)}]{Ingram2019}
{Ingram}, A. 2019, \mnras, 489, 3927

\bibitem[{{Ingram} {et~al.}(2009){Ingram}, {Done}, \& {Fragile}}]{Ingram2009}
{Ingram}, A., {Done}, C., \& {Fragile}, P.~C. 2009, \mnras, 397, L101

\bibitem[{Ingram \& Klis(2013)}]{Ingram2013}
Ingram, A. \& Klis, M. v.~d. 2013, Monthly Notices of the Royal Astronomical
  Society, 434, 1476

\bibitem[{{Ingram} \& {Motta}(2019)}]{IngramMotta2019}
{Ingram}, A.~R. \& {Motta}, S.~E. 2019, \nar, 85, 101524

\bibitem[{{Ji} {et~al.}(2003){Ji}, {Zhang}, {Qu}, \& {Li}}]{Ji2003}
{Ji}, J.~F., {Zhang}, S.~N., {Qu}, J.~L., \& {Li}, T.~P. 2003, \apjl, 584, L23

\bibitem[{{Kara} {et~al.}(2019){Kara}, {Steiner}, {Fabian}, {Cackett},
  {Uttley}, {Remillard}, {Gendreau}, {Arzoumanian}, {Altamirano}, {Eikenberry},
  {Enoto}, {Homan}, {Neilsen}, \& {Stevens}}]{Kara2019}
{Kara}, E., {Steiner}, J.~F., {Fabian}, A.~C., {et~al.} 2019, \nat, 565, 198

\bibitem[{{Karpouzas} {et~al.}(2020){Karpouzas}, {M{\'e}ndez}, {Ribeiro},
  {Altamirano}, {Blaes}, \& {Garc{\'\i}a}}]{Karpouzas2020}
{Karpouzas}, K., {M{\'e}ndez}, M., {Ribeiro}, E.~M., {et~al.} 2020, \mnras,
  492, 1399

\bibitem[{{Kawamura} {et~al.}(2023){Kawamura}, {Done}, {Axelsson}, \&
  {Takahashi}}]{Kawamura2023}
{Kawamura}, T., {Done}, C., {Axelsson}, M., \& {Takahashi}, T. 2023, \mnras,
  519, 4434

\bibitem[{{Kawamuro} {et~al.}(2018){Kawamuro}, {Negoro}, {Yoneyama}, {Ueno},
  {Tomida}, {Ishikawa}, {Sugawara}, {Isobe}, {Shimomukai}, {Mihara},
  {Sugizaki}, {Nakahira}, {Iwakiri}, {Yatabe}, {Takao}, {Matsuoka}, {Kawai},
  {Sugita}, {Yoshii}, {Tachibana}, {Harita}, {Morita}, {Yoshida}, {Sakamoto},
  {Serino}, {Kawakubo}, {Kitaoka}, {Hashimoto}, {Tsunemi}, {Nakajima},
  {Kawase}, {Sakamaki}, {Maruyama}, {Ueda}, {Hori}, {Tanimoto}, {Oda},
  {Morita}, {Yamada}, {Tsuboi}, {Nakamura}, {Sasaki}, {Kawai}, {Sato},
  {Yamauchi}, {Hanyu}, {Hidaka}, {Yamaoka}, \& {Shidatsu}}]{Kawamuro2018}
{Kawamuro}, T., {Negoro}, H., {Yoneyama}, T., {et~al.} 2018, The Astronomer's
  Telegram, 11399, 1

\bibitem[{{K{\"o}nig} {et~al.}(2024){K{\"o}nig}, {Mastroserio}, {Dauser},
  {M{\'e}ndez}, {Wang}, {Garc{\'\i}a}, {Steiner}, {Pottschmidt}, {Ballhausen},
  {Connors}, {Garc{\'\i}a}, {Grinberg}, {Horn}, {Ingram}, {Kara}, {Kallman},
  {Lucchini}, {Nathan}, {Nowak}, {Thalhammer}, {van der Klis}, \&
  {Wilms}}]{Konig2024}
{K{\"o}nig}, O., {Mastroserio}, G., {Dauser}, T., {et~al.} 2024, \aap, 687,
  A284

\bibitem[{{Lyubarskii}(1997)}]{Lyubarskii1997}
{Lyubarskii}, Y.~E. 1997, \mnras, 292, 679

\bibitem[{{Ma} {et~al.}(2023){Ma}, {M{\'e}ndez}, {Garc{\'\i}a}, {Sai}, {Zhang},
  \& {Zhang}}]{Ma2023}
{Ma}, R., {M{\'e}ndez}, M., {Garc{\'\i}a}, F., {et~al.} 2023, \mnras, 525, 854

\bibitem[{{Mastroserio} {et~al.}(2019){Mastroserio}, {Ingram}, \& {van der
  Klis}}]{Mastroserio2019}
{Mastroserio}, G., {Ingram}, A., \& {van der Klis}, M. 2019, \mnras, 488, 348

\bibitem[{{Matsuoka} {et~al.}(2009){Matsuoka}, {Kawasaki}, {Ueno}, {Tomida},
  {Kohama}, {Suzuki}, {Adachi}, {Ishikawa}, {Mihara}, {Sugizaki}, {Isobe},
  {Nakagawa}, {Tsunemi}, {Miyata}, {Kawai}, {Kataoka}, {Morii}, {Yoshida},
  {Negoro}, {Nakajima}, {Ueda}, {Chujo}, {Yamaoka}, {Yamazaki}, {Nakahira},
  {You}, {Ishiwata}, {Miyoshi}, {Eguchi}, {Hiroi}, {Katayama}, \&
  {Ebisawa}}]{Matsuoka2009}
{Matsuoka}, M., {Kawasaki}, K., {Ueno}, S., {et~al.} 2009, \pasj, 61, 999

\bibitem[{{M{\'e}ndez} {et~al.}(2013){M{\'e}ndez}, {Altamirano}, {Belloni}, \&
  {Sanna}}]{Mendez2013}
{M{\'e}ndez}, M., {Altamirano}, D., {Belloni}, T., \& {Sanna}, A. 2013, \mnras,
  435, 2132

\bibitem[{{M{\'e}ndez} {et~al.}(2024){M{\'e}ndez}, {Peirano}, {Garc{\'\i}a},
  {Belloni}, {Altamirano}, \& {Alabarta}}]{Mendez2024}
{M{\'e}ndez}, M., {Peirano}, V., {Garc{\'\i}a}, F., {et~al.} 2024, \mnras, 527,
  9405

\bibitem[{{M{\'e}ndez} \& {van der Klis}(1997)}]{Mendez1997}
{M{\'e}ndez}, M. \& {van der Klis}, M. 1997, \apj, 479, 926

\bibitem[{{Motta} {et~al.}(2012){Motta}, {Homan}, {Mu{\~n}oz Darias},
  {Casella}, {Belloni}, {Hiemstra}, \& {M{\'e}ndez}}]{Motta2012}
{Motta}, S., {Homan}, J., {Mu{\~n}oz Darias}, T., {et~al.} 2012, \mnras, 427,
  595

\bibitem[{{Motta} {et~al.}(2011){Motta}, {Mu{\~n}oz-Darias}, {Casella},
  {Belloni}, \& {Homan}}]{Motta2011}
{Motta}, S., {Mu{\~n}oz-Darias}, T., {Casella}, P., {Belloni}, T., \& {Homan},
  J. 2011, \mnras, 418, 2292

\bibitem[{{Motta}(2016)}]{Motta2016}
{Motta}, S.~E. 2016, Astronomische Nachrichten, 337, 398

\bibitem[{{Mu{\~n}oz-Darias} {et~al.}(2011){Mu{\~n}oz-Darias}, {Motta},
  {Stiele}, \& {Belloni}}]{Munoz2011}
{Mu{\~n}oz-Darias}, T., {Motta}, S., {Stiele}, H., \& {Belloni}, T.~M. 2011,
  \mnras, 415, 292

\bibitem[{{Nowak}(2000)}]{Nowak2000}
{Nowak}, M.~A. 2000, \mnras, 318, 361

\bibitem[{{Nowak} {et~al.}(1999{\natexlab{a}}){Nowak}, {Vaughan}, {Wilms},
  {Dove}, \& {Begelman}}]{Nowak1999}
{Nowak}, M.~A., {Vaughan}, B.~A., {Wilms}, J., {Dove}, J.~B., \& {Begelman},
  M.~C. 1999{\natexlab{a}}, \apj, 510, 874

\bibitem[{Nowak {et~al.}(1997)Nowak, Wagoner, Begelman, \& Lehr}]{Nowak1997}
Nowak, M.~A., Wagoner, R.~V., Begelman, M.~C., \& Lehr, D.~E. 1997, The
  Astrophysical Journal, 477, L91

\bibitem[{{Nowak} {et~al.}(1999{\natexlab{b}}){Nowak}, {Wilms}, \&
  {Dove}}]{NowakNWD}
{Nowak}, M.~A., {Wilms}, J., \& {Dove}, J.~B. 1999{\natexlab{b}}, \apj, 517,
  355

\bibitem[{{Peng} {et~al.}(2023){Peng}, {Zhang}, {Chen}, {Kong}, {Wang},
  {Zhang}, {Ji}, {Tao}, {Qu}, {Ge}, {Shui}, {Li}, {Chang}, {Li}, \&
  {Xiao}}]{Peng2023}
{Peng}, J.~Q., {Zhang}, S., {Chen}, Y.~P., {et~al.} 2023, \mnras, 518, 2521

\bibitem[{{Remillard} \& {McClintock}(2006)}]{Remillard2006}
{Remillard}, R.~A. \& {McClintock}, J.~E. 2006, \araa, 44, 49

\bibitem[{{Remillard} {et~al.}(2002){Remillard}, {Sobczak}, {Muno}, \&
  {McClintock}}]{Remillard2002}
{Remillard}, R.~A., {Sobczak}, G.~J., {Muno}, M.~P., \& {McClintock}, J.~E.
  2002, \apj, 564, 962

\bibitem[{{Reynolds} {et~al.}(1999){Reynolds}, {Young}, {Begelman}, \&
  {Fabian}}]{Reynolds1999}
{Reynolds}, C.~S., {Young}, A.~J., {Begelman}, M.~C., \& {Fabian}, A.~C. 1999,
  \apj, 514, 164

\bibitem[{{Russell} {et~al.}(2019){Russell}, {Baglio}, \&
  {Lewis}}]{Russell2019}
{Russell}, D.~M., {Baglio}, M.~C., \& {Lewis}, F. 2019, The Astronomer's
  Telegram, 12534, 1

\bibitem[{{Shidatsu} {et~al.}(2019){Shidatsu}, {Nakahira}, {Murata}, {Adachi},
  {Kawai}, {Ueda}, \& {Negoro}}]{Shidatsu2019}
{Shidatsu}, M., {Nakahira}, S., {Murata}, K.~L., {et~al.} 2019, \apj, 874, 183

\bibitem[{{Ulowetz} {et~al.}(2019){Ulowetz}, {Myers}, \&
  {Patterson}}]{Ulowetz2019}
{Ulowetz}, J., {Myers}, G., \& {Patterson}, J. 2019, The Astronomer's Telegram,
  12567, 1

\bibitem[{{Uttley} {et~al.}(2014){Uttley}, {Cackett}, {Fabian}, {Kara}, \&
  {Wilkins}}]{Uttley2014}
{Uttley}, P., {Cackett}, E.~M., {Fabian}, A.~C., {Kara}, E., \& {Wilkins},
  D.~R. 2014, \aapr, 22, 72

\bibitem[{{Uttley} {et~al.}(2005){Uttley}, {McHardy}, \&
  {Vaughan}}]{Uttley2005}
{Uttley}, P., {McHardy}, I.~M., \& {Vaughan}, S. 2005, \mnras, 359, 345

\bibitem[{{van der Klis}(1989)}]{vanderklis1989}
{van der Klis}, M. 1989, \araa, 27, 517

\bibitem[{{van der Klis}(1994)}]{vanderklis1994}
{van der Klis}, M. 1994, \apjs, 92, 511

\bibitem[{{van der Klis}(2006)}]{vanderKlis2006}
{van der Klis}, M. 2006, in Compact stellar X-ray sources, Vol.~39, 39--112

\bibitem[{{van der Klis} {et~al.}(1987){van der Klis}, {Hasinger}, {Stella},
  {Langmeier}, {van Paradijs}, \& {Lewin}}]{vanderklis1987}
{van der Klis}, M., {Hasinger}, G., {Stella}, L., {et~al.} 1987, \apjl, 319,
  L13

\bibitem[{{Vaughan} \& {Nowak}(1997)}]{Vaughan1997}
{Vaughan}, B.~A. \& {Nowak}, M.~A. 1997, \apjl, 474, L43

\bibitem[{{Vaughan} {et~al.}(1997){Vaughan}, {van der Klis}, {M{\'e}ndez}, {van
  Paradijs}, {Wijnands}, {Lewin}, {Lamb}, {Psaltis}, {Kuulkers}, \&
  {Oosterbroek}}]{Vaughan1998}
{Vaughan}, B.~A., {van der Klis}, M., {M{\'e}ndez}, M., {et~al.} 1997, \apjl,
  483, L115

\bibitem[{{Veledina}(2016)}]{Veledina2016}
{Veledina}, A. 2016, \apj, 832, 181

\bibitem[{{Wijnands} {et~al.}(1999){Wijnands}, {Homan}, \& {van der
  Klis}}]{Wijnands1999}
{Wijnands}, R., {Homan}, J., \& {van der Klis}, M. 1999, \apjl, 526, L33

\bibitem[{{Zhou} {et~al.}(2022){Zhou}, {Zhang}, {Song}, {Qu}, {Zhang}, {Ma},
  {Tuo}, {Ge}, {Wang}, {Zhang}, \& {Tao}}]{Zhou2022}
{Zhou}, D.-K., {Zhang}, S.-N., {Song}, L.-M., {et~al.} 2022, \mnras, 515, 1914

\end{thebibliography}

\newpage

\appendix
\section{Best-fitting model for each observation of the HSS-to-LHS transition}
\label{sec:appendix}

\begin{figure*}
    \centering
    \includegraphics[width=\columnwidth]{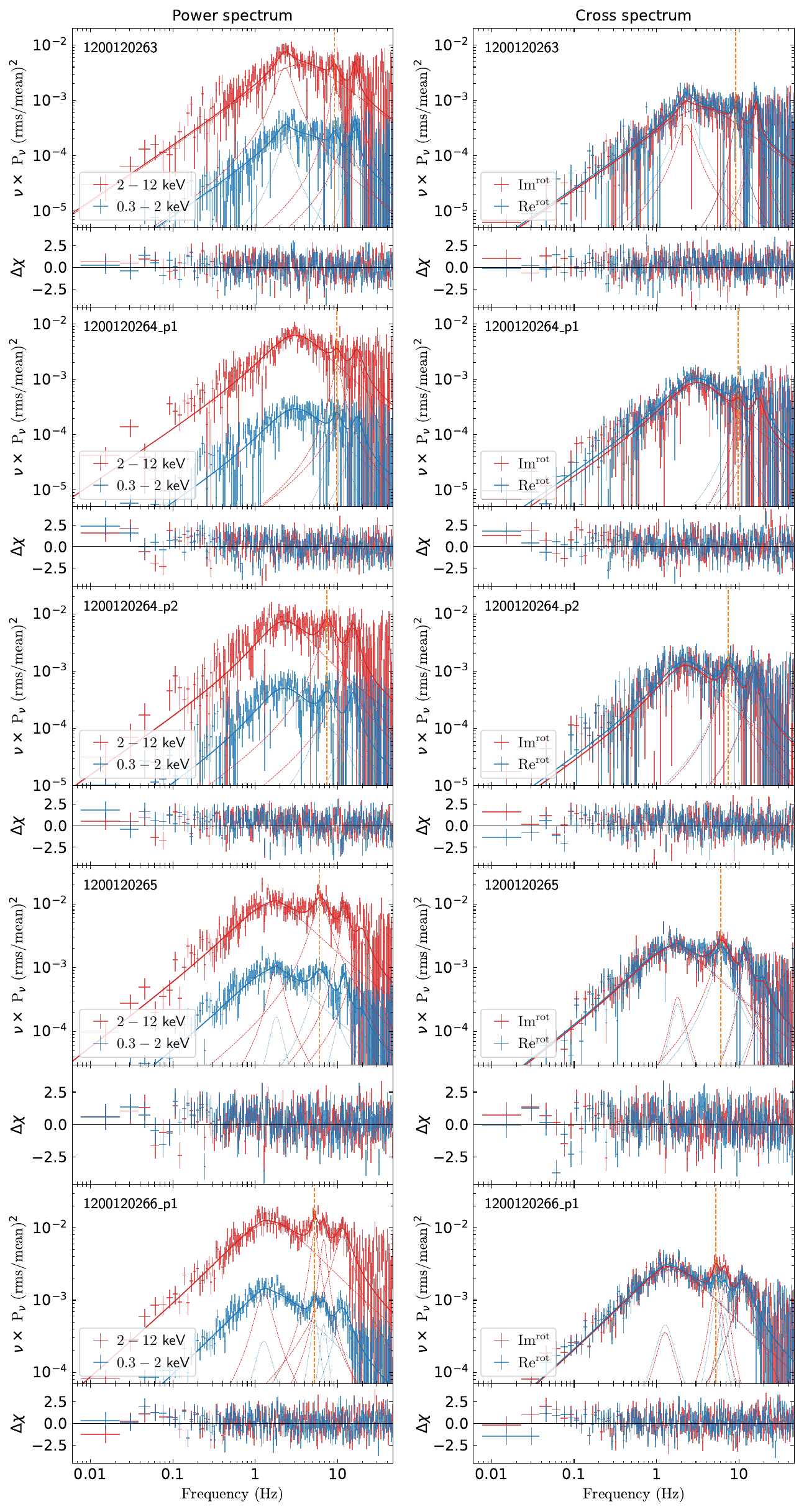}
    \includegraphics[width=\columnwidth]{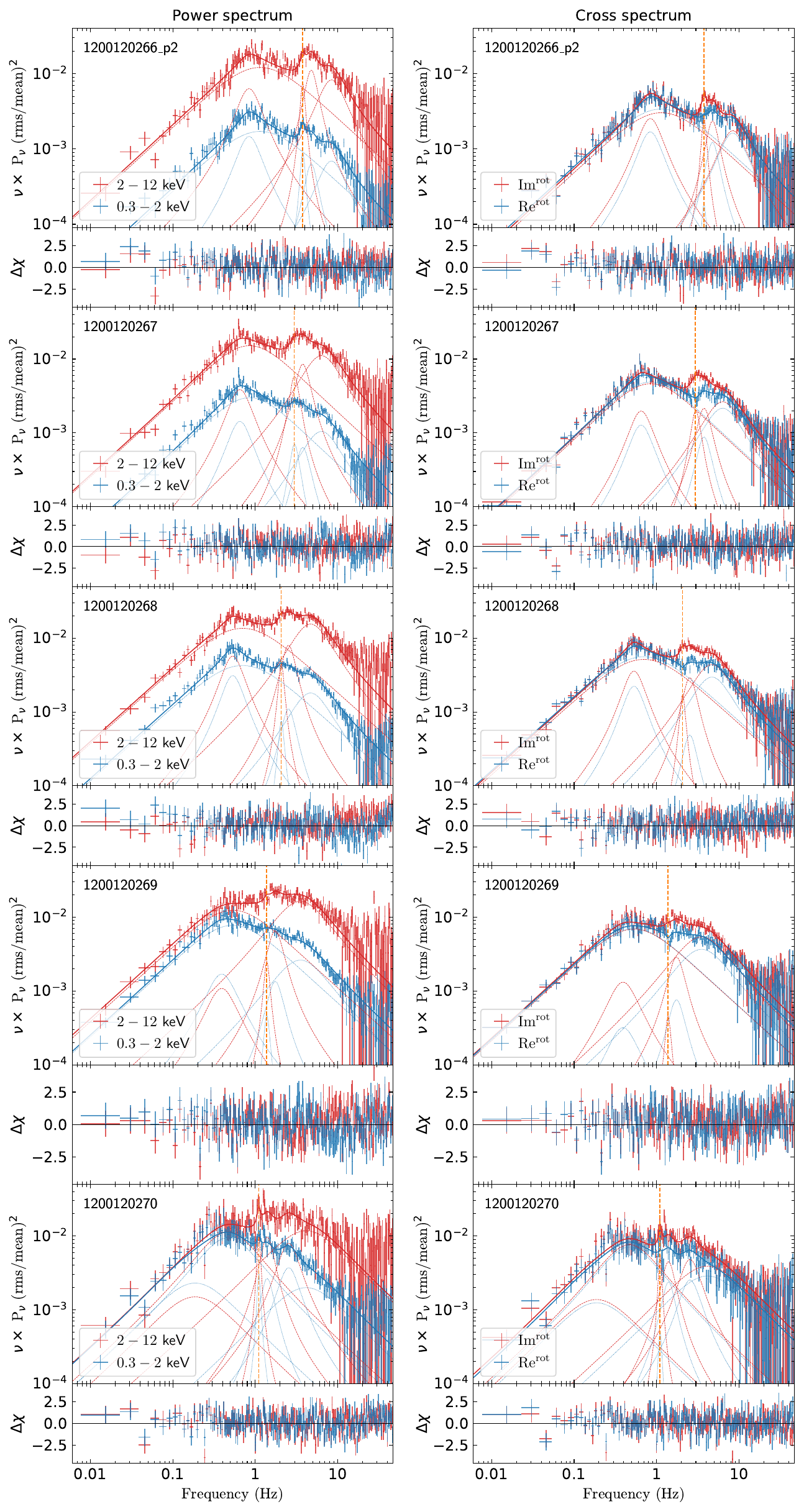}
    \caption{First and third columns: PS in two energy bands of \NICER observations of \MAXI covering the HSS-to-LHS transition. The $0.3-2$~keV data is shown in blue while the $2-12$~keV one is in red, both with the best-fitting model (solid line) consisting of a combination of Lorentzian functions (dotted lines). Second and fourth columns: Real and Imaginary parts of the CS between the same two energy bands rotated by $45^{\circ}$. We plot $\rm{Re}~cos(\pi/4) -\rm{Im}~sin(\pi/4)$ in blue and $\rm{Re}~sin(\pi/4) +\rm{Im}~cos(\pi/4)$ in red, with the best-fitting model assuming the constant phase-lag model. In each panel, residuals with respect to the model, defined as $\Delta \chi = {\rm (data-model)/error}$, are also plotted. The orange dashed vertical line in each panel corresponds to the centroid frequency of the imaginary/real QPO in the model.}
    \label{fig:allfits}
\end{figure*}

We fit the  \NICER observations of \MAXI covering the HSS-to-LHS transition; we depict those observations with light blue diamonds and blue dots in Fig.~\ref{fig:HID}. We use a multi-Lorentzian model to fit simultaneously the PS in the $0.3-2$~keV and $2-12$~keV energy bands and the Real and Imaginary parts of the CS between these two energy bands. We employed the technique introduced by \cite{Mendez2024} assuming the constant phase-lag model. In Fig.~\ref{fig:allfits}, we show the fit of the power spectra in the two bands (first and third columns) and the real and imaginary parts of the CS (second and fourth columns) for each observation of the soft-to-hard transition, with their corresponding residuals. In Table \ref{tab:longtable} we present the best-fitting values for each parameter, along with the corresponding $1-\sigma$ uncertainties.  The rms amplitude at each energy band is the square root of the Lorentzian normalisation and the phase lag is the argument of the cosine and the sine functions that multiply the Lorentzian in the Real and Imaginary parts of the CS, respectively.

\begin{table*}
\caption{Best-fitting parameters of the \NICER observations of \MAXI covering the HSS-to-LHS transition, and their corresponding $1-\sigma$ uncertainties.}
\centering
\small
\scalebox{0.97}{
\begin{tabular}{cccccccc}
\hline
\noalign{\smallskip}
Observation & Component & $\nu_0$ & FWHM & ${{\rm rms}}_{{0.3-2~{{\rm keV}}}}$  & ${{\rm rms}}_{{2-12~{{\rm keV}}}}$ & Phase lags &  $\chi^2/{\rm dof}$ \\ 
 & & (Hz) & (Hz) & (\%) & (\%) & (rad) &  \\ \hline
\noalign{\smallskip} 
1200120263 & & & & & & & $835/828$ \\ 
\phantom{obsID} & Lorentzian 1 & 2.22 $\pm$ 0.05 & 1.0 $\pm$ 0.3 & 1.0 $\pm$ 0.2 & 5.0 $\pm$ 0.7 & -0.31 $\pm$ 0.13 \\ 
\phantom{obsID} & Lorentzian 2 & 1.51 $\pm$ 0.28 & 6.4 $\pm$ 0.9 & 2.4 $\pm$ 0.1 & 10.6 $\pm$ 0.3 & -0.02 $\pm$ 0.04 \\ 
\phantom{obsID} & Lorentzian 3 & 9.14 $\pm$ 0.29 & 2.1 $\pm$ 0.6 & 0.7 $\pm$ 0.1 & 2.7 $\pm$ 0.5 & 0.01 $\pm$ 0.20 \\ 
\phantom{obsID} & Lorentzian 4 & 16.16 $\pm$ 0.20 & 2.5 $\pm$ 0.5 & 1.0 $\pm$ 0.1 & 3.0 $\pm$ 0.4 & -0.16 $\pm$ 0.14 \\ 
\hline 
\noalign{\smallskip} 
1200120264\_p1 & & & & & & & $937/833$ \\ 
\phantom{obsID} & Lorentzian 1 & 2.17 $\pm$ 0.06 & 4.4 $\pm$ 0.2 & 2.4 $\pm$ 0.0 & 11.1 $\pm$ 0.2 & -0.09 $\pm$ 0.02 \\ 
\phantom{obsID} & Lorentzian 2 & 9.72 $\pm$ 0.29 & 2.8 $\pm$ 0.7 & 0.8 $\pm$ 0.1 & 2.7 $\pm$ 0.4 & -0.39 $\pm$ 0.17 \\ 
\phantom{obsID} & Lorentzian 3 & 16.99 $\pm$ 0.48 & 5.4 $\pm$ 1.4 & 0.9 $\pm$ 0.1 & 3.5 $\pm$ 0.4 & -0.24 $\pm$ 0.14 \\ 
\hline 
\noalign{\smallskip} 
1200120264\_p2 & & & & & & & $936/833$ \\ 
\phantom{obsID} & Lorentzian 1 & 1.74 $\pm$ 0.05 & 2.9 $\pm$ 0.2 & 3.0 $\pm$ 0.1 & 11.4 $\pm$ 0.3 & -0.08 $\pm$ 0.03 \\ 
\phantom{obsID} & Lorentzian 2 & 7.41 $\pm$ 0.14 & 3.0 $\pm$ 0.5 & 1.4 $\pm$ 0.1 & 6.3 $\pm$ 0.5 & -0.06 $\pm$ 0.09 \\ 
\phantom{obsID} & Lorentzian 3 & 15.26 $\pm$ 0.29 & 4.2 $\pm$ 0.7 & 1.2 $\pm$ 0.1 & 4.8 $\pm$ 0.4 & 0.02 $\pm$ 0.13 \\ 
\hline 
\noalign{\smallskip} 
1200120265 & & & & & & & $840/824$ \\ 
\phantom{obsID} & Lorentzian 1 & 1.77 $\pm$ 0.12 & $0.6^\dagger$ & 0.9 $\pm$ 0.3 & 2.4 $\pm$ 1.0 & 0.13 $\pm$ 0.59 \\ 
\phantom{obsID} & Lorentzian 2 & 0.89 $\pm$ 0.05 & 2.7 $\pm$ 0.1 & 4.5 $\pm$ 0.1 & 15.2 $\pm$ 0.3 & -0.03 $\pm$ 0.03 \\ 
\phantom{obsID} & Lorentzian 3 & 6.03 $\pm$ 0.09 & 3.1 $\pm$ 0.3 & 2.2 $\pm$ 0.1 & 8.3 $\pm$ 0.4 & 0.18 $\pm$ 0.05 \\ 
\phantom{obsID} & Lorentzian 4 & 11.94 $\pm$ 0.12 & 2.6 $\pm$ 0.4 & 1.4 $\pm$ 0.1 & 5.1 $\pm$ 0.4 & -0.06 $\pm$ 0.09 \\ 
\phantom{obsID} & Lorentzian 5 & 19.39 $\pm$ 1.48 & 7.3 $\pm$ 4.4 & 0.7 $\pm$ 0.2 & 3.8 $\pm$ 0.9 & -0.15 $\pm$ 0.33 \\ 
\hline 
\noalign{\smallskip} 
1200120266\_p1 & & & & & & & $782/824$ \\ 
\phantom{obsID} & Lorentzian 1 & 1.23 $\pm$ 0.12 & $0.6^\dagger$ & 1.4 $\pm$ 0.3 & 2.9 $\pm$ 1.1 & -0.11 $\pm$ 0.52 \\ 
\phantom{obsID} & Lorentzian 2 & 0.67 $\pm$ 0.07 & 2.7 $\pm$ 0.1 & 5.5 $\pm$ 0.1 & 17.1 $\pm$ 0.3 & -0.02 $\pm$ 0.03 \\ 
\phantom{obsID} & Lorentzian 3 & 5.23 $\pm$ 0.05 & 1.2 $\pm$ 0.2 & 1.4 $\pm$ 0.1 & 5.4 $\pm$ 0.5 & 0.32 $\pm$ 0.08 \\ 
\phantom{obsID} & Lorentzian 4 & 6.72 $\pm$ 0.08 & 1.3 $\pm$ 0.2 & 1.2 $\pm$ 0.1 & 4.5 $\pm$ 0.5 & 0.41 $\pm$ 0.10 \\ 
\phantom{obsID} & Lorentzian 5 & 11.06 $\pm$ 0.15 & 5.6 $\pm$ 0.6 & 2.0 $\pm$ 0.1 & 7.4 $\pm$ 0.3 & -0.01 $\pm$ 0.05 \\ 
\hline 
\noalign{\smallskip} 
1200120266\_p2 & & & & & & & $999/823$ \\ 
\phantom{obsID} & Lorentzian 1 & 0.79 $\pm$ 0.02 & 0.6 $\pm$ 0.1 & 3.9 $\pm$ 0.5 & 8.0 $\pm$ 1.1 & 0.20 $\pm$ 0.09 \\ 
\phantom{obsID} & Lorentzian 2 & 0.10 $\pm$ 0.11 & 2.3 $\pm$ 0.1 & 7.2 $\pm$ 0.3 & 19.2 $\pm$ 0.4 & -0.06 $\pm$ 0.03 \\ 
\phantom{obsID} & Lorentzian 3 & 3.75 $\pm$ 0.03 & 0.5 $\pm$ 0.1 & 1.4 $\pm$ 0.1 & 4.0 $\pm$ 0.6 & 0.72 $\pm$ 0.10 \\ 
\phantom{obsID} & Lorentzian 4 & 4.69 $\pm$ 0.10 & 1.8 $\pm$ 0.2 & 1.9 $\pm$ 0.2 & 7.9 $\pm$ 0.7 & 0.31 $\pm$ 0.06 \\ 
\phantom{obsID} & Lorentzian 5 & 7.71 $\pm$ 0.22 & 6.7 $\pm$ 0.5 & 2.8 $\pm$ 0.1 & 9.7 $\pm$ 0.5 & 0.04 $\pm$ 0.04 \\ 
\hline 
\noalign{\smallskip} 
1200120267 & & & & & & & $971/824$ \\ 
\phantom{obsID} & Lorentzian 1 & 0.61 $\pm$ 0.02 & $0.5^\dagger$ & 3.8 $\pm$ 0.4 & 6.2 $\pm$ 0.9 & 0.21 $\pm$ 0.15 \\ 
\phantom{obsID} & Lorentzian 2 & 0.25 $\pm$ 0.04 & 1.6 $\pm$ 0.1 & 9.1 $\pm$ 0.1 & 20.5 $\pm$ 0.3 & -0.03 $\pm$ 0.02 \\ 
\phantom{obsID} & Lorentzian 3 & 2.96 $\pm$ 0.06 & 0.9 $\pm$ 0.2 & 1.7 $\pm$ 0.4 & 5.0 $\pm$ 1.3 & 0.97 $\pm$ 0.18 \\ 
\phantom{obsID} & Lorentzian 4 & 3.68 $\pm$ 0.16 & 1.6 $\pm$ 0.5 & 2.0 $\pm$ 0.5 & 7.3 $\pm$ 1.8 & 0.40 $\pm$ 0.11 \\ 
\phantom{obsID} & Lorentzian 5 & 5.30 $\pm$ 0.33 & 7.0 $\pm$ 0.3 & 4.0 $\pm$ 0.2 & 13.1 $\pm$ 0.7 & 0.10 $\pm$ 0.03 \\ 
\hline 
\noalign{\smallskip} 
1200120268 & & & & & & & $992/823$ \\ 
\phantom{obsID} & Lorentzian 1 & 0.50 $\pm$ 0.01 & 0.4 $\pm$ 0.1 & 5.4 $\pm$ 0.6 & 7.4 $\pm$ 0.8 & 0.22 $\pm$ 0.09 \\ 
\phantom{obsID} & Lorentzian 2 & 0.02 $\pm$ 0.05 & 1.4 $\pm$ 0.1 & 11.5 $\pm$ 0.3 & 20.5 $\pm$ 0.4 & -0.04 $\pm$ 0.02 \\ 
\phantom{obsID} & Lorentzian 3 & 2.06 $\pm$ 0.03 & 0.3 $\pm$ 0.1 & 1.4 $\pm$ 0.4 & 3.4 $\pm$ 1.0 & 1.16 $\pm$ 0.23 \\ 
\phantom{obsID} & Lorentzian 4 & 2.48 $\pm$ 0.09 & 1.0 $\pm$ 0.2 & 2.6 $\pm$ 0.4 & 7.0 $\pm$ 1.1 & 0.62 $\pm$ 0.08 \\ 
\phantom{obsID} & Lorentzian 5 & 3.87 $\pm$ 0.15 & 5.3 $\pm$ 0.1 & 5.3 $\pm$ 0.2 & 15.6 $\pm$ 0.4 & 0.19 $\pm$ 0.02 \\ 
\hline 
\noalign{\smallskip} 
1200120269 & & & & & & & $922/824$ \\ 
\phantom{obsID} & Lorentzian 1 & 0.35 $\pm$ 0.04 & $0.3^\dagger$ & 4.6 $\pm$ 1.0 & 3.6 $\pm$ 1.4 & 0.55 $\pm$ 0.33 \\ 
\phantom{obsID} & Lorentzian 2 & 0.10 $\pm$ 0.04 & 1.0 $\pm$ 0.1 & 15.0 $\pm$ 0.3 & 19.0 $\pm$ 0.4 & -0.01 $\pm$ 0.03 \\ 
\phantom{obsID} & Lorentzian 3 & 1.37 $\pm$ 0.04 & 0.2 $\pm$ 0.1 & 1.1 $\pm$ 0.3 & 2.5 $\pm$ 0.8 & 1.93 $\pm$ 0.37 \\ 
\phantom{obsID} & Lorentzian 4 & 1.70 $\pm$ 0.01 & 0.8 $\pm$ 0.2 & 3.0 $\pm$ 0.4 & 6.7 $\pm$ 1.0 & 0.53 $\pm$ 0.12 \\ 
\phantom{obsID} & Lorentzian 5 & 2.39 $\pm$ 0.16 & 5.2 $\pm$ 0.1 & 7.3 $\pm$ 0.3 & 18.2 $\pm$ 0.5 & 0.19 $\pm$ 0.02 \\ 
\hline 
\noalign{\smallskip} 
1200120270 & & & & & & & $860/819$ \\ 
\phantom{obsID} & Lorentzian 1 & 0.08 $\pm$ 0.06 & $0.3^\dagger$ & 7.7 $\pm$ 1.8 & 6.2 $\pm$ 2.5 & 0.05 $\pm$ 0.24 \\ 
\phantom{obsID} & Lorentzian 2 & 0.30 $\pm$ 0.06 & 0.7 $\pm$ 0.1 & 14.2 $\pm$ 1.2 & 15.5 $\pm$ 1.5 & 0.04 $\pm$ 0.05 \\ 
\phantom{obsID} & Lorentzian 3 & 1.10 $\pm$ 0.01 & 0.1 $\pm$ 0.1 & 2.1 $\pm$ 0.4 & 3.9 $\pm$ 0.7 & 0.70 $\pm$ 0.26 \\ 
\phantom{obsID} & Lorentzian 4 & 1.37 $\pm$ 0.05 & 0.5 $\pm$ 0.1 & 3.9 $\pm$ 0.7 & 7.5 $\pm$ 1.2 & 0.30 $\pm$ 0.12 \\ 
\phantom{obsID} & Lorentzian 5 & 2.42 $\pm$ 0.09 & 1.6 $\pm$ 0.4 & 5.7 $\pm$ 0.9 & 8.4 $\pm$ 2.1 & 0.30 $\pm$ 0.09 \\ 
\phantom{obsID} & Lorentzian 6 & 1.82 $\pm$ 0.85 & 7.5 $\pm$ 1.1 & 7.1 $\pm$ 1.0 & 17.2 $\pm$ 1.8 & 0.15 $\pm$ 0.04 \\ 
\hline 
\end{tabular}}
\flushleft{$\dagger$ Parameter value frozen during the fitting.}
\tablefoot{We simultaneously fit a multi-Lorentzian model to the PS in two energy bands, $0.3-2$~keV and $2-12$~keV, and the CS between these two energy bands, following the technique introduced by \cite{Mendez2024}, assuming the constant phase-lag model.}
\label{tab:longtable}
\end{table*}

\section{ rms and phase-lag energy-spectra}
\label{appendixrmsylags}

\begin{figure*}
    \centering
    \includegraphics[width=0.58\textwidth]{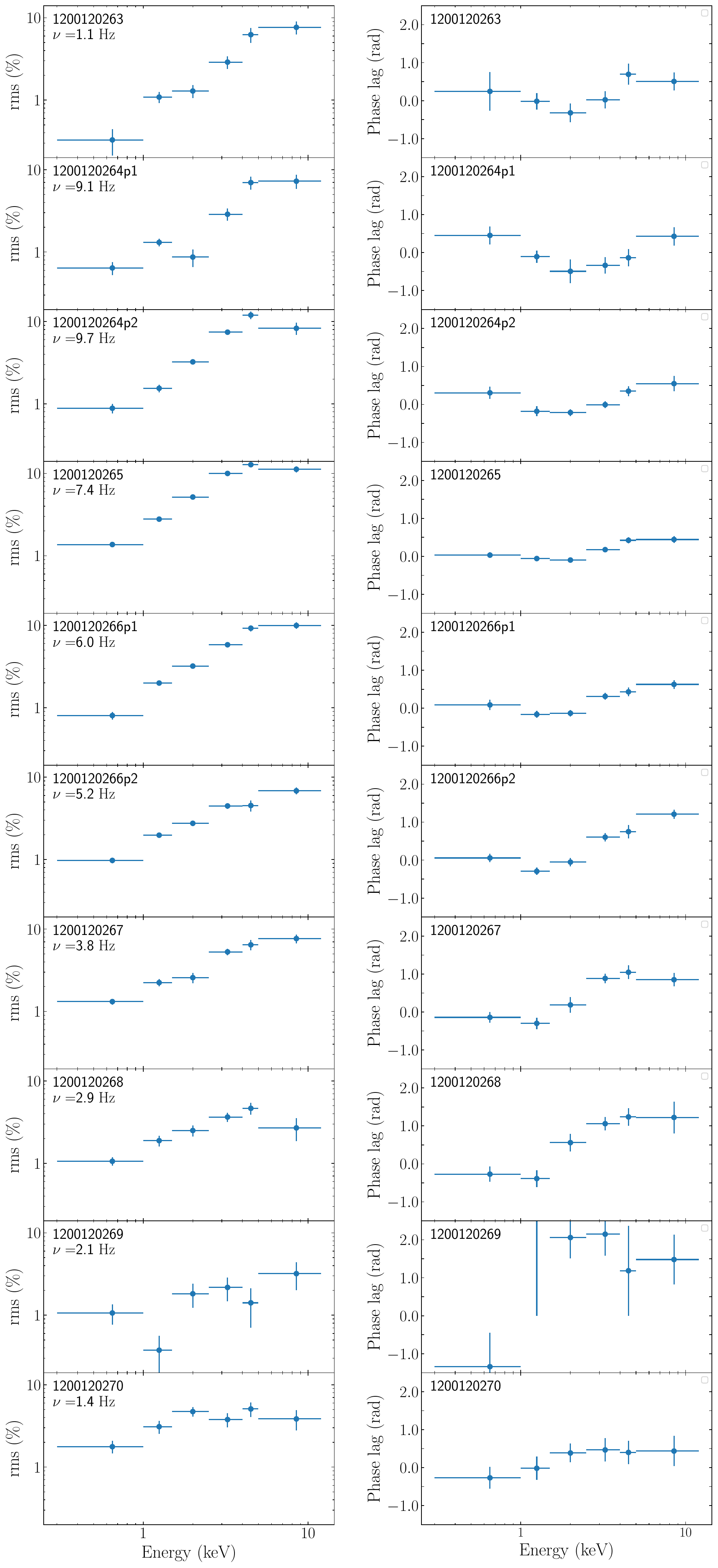}

    \caption{The rms (left panels) and phase-lag (right panels) energy spectra of the imaginary/real QPO for the \NICER observations of \MAXI in the HSS-to-LHS transition. Horizontal lines cover the range of the energy band corresponding to each marker.}
    \label{fig:allrmsylags}
\end{figure*}

In Fig.~\ref{fig:allrmsylags}, we show the rms and phase-lag energy spectra of the real/imaginary QPO for all the observations in the HSS-to-LHS transition of \MAXI. To produce these spectra, we extract the PS in six energy bands, $0.3-1.0$, $1.0-1.5$, $1.5-2.5$, $2.5-4.0$, $4.0-5.0$, $5.0-12.0$~keV, and we fit each of them simultaneously with the PS in the full energy band, $0.3-12$~keV, and the CS between each narrow band and the full energy band. For each observation, we use the corresponding best-fitting model presented in Appendix~\ref{sec:appendix}, with the centroid frequency and FWHM of every Lorentzian component fixed. We use the full energy band as the reference band for the phase-lags. 

\end{document}